\documentclass[useAMS,usenatbib]{mn2e}
\usepackage{graphicx}
\usepackage{rotating}
\usepackage{aas_macros}
\usepackage{amsmath}
\usepackage{color}

\def\Re{{R$_{\rm e}$}}
\def\etal{{\it et al. }} 

\title[AIMSS I] {The AIMSS Project I: Bridging the Star Cluster - Galaxy Divide\thanks{Based on observations obtained at the Southern 
Astrophysical Research (SOAR) telescope, which is a joint project of the Minist\'{e}rio da Ci\^{e}ncia, Tecnologia, e Inova\c{c}\~{a}o 
(MCTI) da Rep\'{u}blica Federativa do Brasil, the U.S. National Optical Astronomy Observatory (NOAO), the University of North 
Carolina at Chapel Hill (UNC), and Michigan State University (MSU).} 
\thanks{Based on observations made with the Southern African Large Telescope (SALT).}
\thanks{Some of the data presented herein were obtained at the W.M. Keck Observatory, which is operated as a scientific partnership 
among the California Institute of Technology, the University of California and the National Aeronautics and Space Administration.
The Observatory was made possible by the generous financial support of the W.M. Keck Foundation.}
\thanks{Based on observations made with the (name of the telescope) operated on the island of La Palma by the Isaac Newton Group 
in the Spanish Observatorio del Roque de los Muchachos of the Instituto de Astrof'sica de Canarias.}
\thanks{Based on observations obtained at the Gemini Observatory, which is operated by the 
Association of Universities for Research in Astronomy, Inc., under a cooperative agreement 
with the NSF on behalf of the Gemini partnership: the National Science Foundation 
(United States), the National Research Council (Canada), CONICYT (Chile), the Australian 
Research Council (Australia), Minist\'{e}rio da Ci\^{e}ncia, Tecnologia e Inova\c{c}\~{a}o 
(Brazil) and Ministerio de Ciencia, Tecnolog\'{i}a e Innovaci\'{o}n Productiva (Argentina).}} 

\author[Norris \etal] {Mark A. Norris$^{1,2}$\thanks{norris@mpia.de}, Sheila J. Kannappan$^{2}$, Duncan A. Forbes$^{3}$, Aaron J. Romanowsky$^{4,5}$, \newauthor 
 Jean P. Brodie$^{5}$, Favio Ra\'ul Faifer$^{6,7}$, Avon Huxor$^{8}$, Claudia Maraston$^{9}$, \newauthor 
 Amanda J. Moffett$^{2}$, Samantha J. Penny$^{10}$, Vincenzo Pota$^{3}$, Anal\'ia Smith-Castelli$^{6,7}$, \newauthor
 Jay Strader$^{11}$, David Bradley$^{2}$, Kathleen D. Eckert$^{2}$, Dora Fohring$^{12,13}$, JoEllen McBride$^{2}$, \newauthor
 David V. Stark$^{2}$, Ovidiu Vaduvescu$^{12}$ \\
 \\
 \\
  $^1$ Max Planck Institut f\"{u}r Astronomie, K\"{o}nigstuhl 17, D-69117, Heidelberg, Germany \\
  $^2$ Dept. of Physics and Astronomy UNC-Chapel Hill, CB 3255, Phillips Hall, Chapel Hill, NC 27599-3255, USA \\
  $^3$ Centre for Astrophysics \& Supercomputing, Swinburne University, Hawthorn, VIC 3122, Australia\\ 
  $^4$ Department of Physics and Astronomy, San Jos\'{e} State University, One Washington Square, San Jose, CA 95192, USA\\
  $^5$ University of California Observatories, 1156 High Street, Santa Cruz, CA 95064, USA \\
  $^6$ Facultad de Ciencias Astron\'{o}micas y Geof\'{i}sicas, Universidad Nacional de La Plata, Paseo del Bosque, B1900FWA, La Plata, Argentina \\
  $^7$ Instituto de Astrof\'{i}sica de La Plata (CCT-La Plata, CONICET-UNLP), Paseo del Bosque, B1900FWA, La Plata, Argentina \\
  $^8$ Astronomisches Rechen-Institut, Zentrum f\"{u}r Astronomie der Universit\"{a}t Heidelberg, M\"{o}nchstra{\ss}e 12-14, D-69120 Heidelberg, Germany \\
  $^9$ Institute of Cosmology and Gravitation, Dennis Sciama Building, Burnaby Road, Portsmouth PO1 3FX \\
  $^{10}$ School of Physics, Monash University, Clayton, Victoria 3800, Australia \\
  $^{11}$ Department of Physics and Astronomy, Michigan State University, East Lansing, Michigan 48824, USA\\
  $^{12}$ Isaac Newton Group of Telescopes, Apto. 321, E-38700 Santa Cruz de la Palma, Canary Islands, Spain \\
  $^{13}$ Department of Physics, Centre for Advanced Instrumentation, University of Durham, South Road, Durham DH1 3LE, United Kingdom
  }
\begin{document}

\date{Accepted 2014 ***. Received 2014 ***; in original form ***}

\pagerange{\pageref{firstpage}--\pageref{lastpage}} \pubyear{2014}

\maketitle

\label{firstpage}

\begin{abstract}
We describe the structural and kinematic properties of the first compact stellar systems 
discovered by the AIMSS project. These spectroscopically confirmed objects have sizes ($\sim$6 $<$ R$_{\rm e}$ 
[pc] $<$ 500) and masses ($\sim$2$\times$10$^{6}$ $<$ M$_*$/M$_\odot$ $<$ 6$\times$10$^{9}$) spanning 
the range of massive globular clusters (GCs), ultra compact dwarfs (UCDs) and compact elliptical galaxies (cEs),
completely filling the gap between star clusters and galaxies.

Several objects are close analogues to the prototypical cE, M32. These objects, which 
are more massive than previously discovered UCDs of the same size, further call into question 
the existence of a tight mass--size trend for compact stellar systems, while simultaneously strengthening 
the case for a universal ``zone of avoidance" for dynamically hot stellar systems in the mass--size plane.

Overall, we argue that there are two classes of compact stellar systems: 1) massive star 
clusters and 2) a population closely related to galaxies. Our data provide indications for a further division 
of the galaxy-type UCD/cE population into two groups, one population that we associate with objects formed by the stripping 
of nucleated dwarf galaxies, and a second population that formed through the stripping of bulged galaxies or 
are lower-mass analogues of classical ellipticals. We find compact stellar systems around 
galaxies in low to high density environments, demonstrating that the physical processes responsible for forming 
them do not only operate in the densest clusters.

\end{abstract}


\section{Introduction}

In the past 15 years there has been a revolution in the study of low mass stellar systems. It 
began with the discovery \citep{Hilker99,Drinkwater00} in the Fornax cluster of a population of 
generally old and compact objects with luminosity/mass and size intermediate between those 
of globular clusters (GCs) and the few then-known compact elliptical galaxies (cEs). These 
objects, known as ultra compact dwarfs \citep[UCDs:][]{Phillips01}, became the first in a series 
of stellar systems found to exist with properties between star clusters and galaxies. They 
were followed by a zoo of objects inhabiting slightly different regions of the size--luminosity 
parameter space. These new objects included extended star clusters such as ``Faint Fuzzies"
\citep{Larsen&BrodieFF1,Larsen02} and ``Extended Clusters" \citep{Huxor05,Huxor11,Brodie11,
Forbes13}, additional MW and M31 dwarf spheroidals and ultra-faint dwarf galaxies 
\citep[e.g.][]{Willman05a,Zucker06,Zucker07,Belokurov07}, and a host of new cEs 
\citep[][]{Mieske05,Chilingarian07,SmithCastelli08,Chilingarian09,Price09} that fill the gap 
between M32 and ``normal" elliptical galaxies.

These discoveries have broken the simple division thought to exist between star clusters and 
galaxies, with UCDs displaying properties that lie between those of ``classical" GCs and early-type 
galaxies. Naturally this situation has led to a search for unifying scaling relations, and therefore 
formation scenarios, for the various compact stellar systems (CSSs) and early-type galaxy populations.

Initial indications of a tight mass--size relation for the UCD and cE populations that extend from the 
massive GC (i.e. stellar mass $>$ 2$\times$10$^{6}$M$_\odot$) to elliptical galaxy regime 
\citep{Hasegan05,Kissler-Patig06,Dabringhausen08,Murray09,Norris&Kannappan11,Misgeld&Hilker11} 
have been called into question by the discovery of extended but faint star clusters that broaden the 
previously observed tight relation for UCDs at fainter magnitudes \citep[see e.g.][]{Brodie11,Forbes13}.
Investigating the reality of such trends requires a more systematic and homogeneous sample of CSSs 
than currently exists. 

In this paper series we present the archive of intermediate mass stellar systems (AIMSS) survey.
The goal of this survey is to produce a comprehensive catalog of spectroscopically confirmed 
CSSs of all types which have resolved sizes from Hubble Space Telescope (HST) photometry, as well
as homogeneous stellar mass estimates, spectroscopically determined velocity dispersions, and stellar 
population information. This catalog will then be used to systematically investigate the formation
of CSSs and their relationships with other stellar systems. 

In order to achieve this goal we have undertaken a search of all available archival HST images to 
find CSS candidates. We have deliberately broadened the selection limits traditionally used to find 
CSSs, both to probe the limits of CSS formation and to avoid producing spurious trends in CSS properties. 
One of the first results of the AIMSS survey presented in this paper is the discovery of further examples 
of a class of extremely dense stellar systems which broaden the previously suggested 
mass/luminosity--size trend to \textit{brighter} magnitudes. 

The AIMSS survey also includes a key additional parameter -- central velocity dispersion. The central 
velocity dispersion of stars has been shown to be one of the best predictors of galaxy properties
\citep[e.g. ][]{Forbes99,Cappellari06,Graves09}. It can also provide clues to the evolutionary history of 
a galaxy since, for example, tidal stripping will tend to reduce both the size and luminosity of a galaxy 
but its velocity dispersion will remain largely unchanged \citep[see e.g.][]{Bender92,Chilingarian09} 
and hence will remain a reliable signature of its past form. 

In fact \citet{Chilingarian09} showed that when their simulated disc galaxy on a circular 
orbit around a cluster potential is stripped severely enough to lose $\sim$75$\%$ of its original mass, 
the global velocity dispersion is essentially unaffected (their Figure 1.). This is because as stripping progresses 
it is increasingly the tightly bound central stellar structure (either nucleus or bulge) that comes to dominate the
global light distribution of the galaxy, and the dispersion of this is relatively unaffected by the loss of an
outer dark matter halo. The central velocity dispersion, which is always dominated by the stellar component
of the galaxy, is likely to be less affected by stripping, at least until the point where the central mass
component itself begins to lose mass.

Although the number of objects with sizes and luminosities 
intermediate between those classified as UCDs and cEs has grown substantially over the last few years, the 
number with measured velocity dispersions has not kept pace. In the compilation of  \citet{Forbes08}, there 
were only two objects shown in their plot of velocity dispersion against luminosity in the gap between 
UCDs and cEs. Here we present velocity dispersions for 20 objects, many of which lie in this gap.

In the current paper we present the catalog of properties, and examine the mass -- size, mass -- surface 
mass density, and mass -- $\sigma$ behaviour of the first 28 (20 of which have velocity dispersion
measurements) of our objects to be spectroscopically confirmed. In future papers in this series we will 
examine the dynamics, stellar populations, and mass-to-light ratios of compact stellar systems (CSSs) 
in more detail. An additional paper will provide the full photometric catalog of candidate compact stellar systems.

\section{Sample}

\subsection{AIMSS Target Selection}

Our experience with the pilot program of this survey \citep{Norris&Kannappan11} demonstrated that 
\textit{inferred} effective radii and absolute magnitudes (both determined by assuming physical 
association between the candidate CSS and an adjacent, larger galaxy), when combined with the 
existence of a hard edge in the luminosity--size distribution of compact stellar systems (as seen in 
\citealt{Misgeld&Hilker11,Norris&Kannappan11,Brodie11}) can be used to reliably select CSSs 
for spectroscopic follow-up. 
 
\begin{figure} 
   \centering
   \begin{turn}{0}
   \includegraphics[scale=1.0]{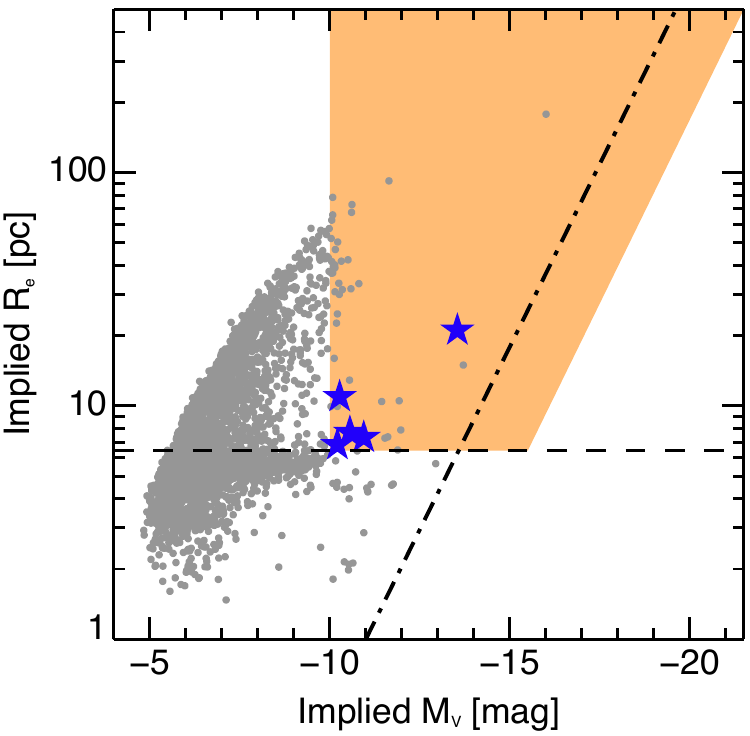}
   \end{turn} 
   \caption{Implied luminosity--size plot for objects detected in an HST ACS pointing centred on NGC~4649,
   created assuming that all objects detected are at the same distance as NGC~4649. 
   The grey dots are all objects detected by SExtractor in the ACS image. The shaded region indicates the 
   selection region, the dashed horizontal line is the HST resolution limit at the distance of NGC~4649. The
   dot-dashed line shows the edge of the ``zone of avoidance" for early-type galaxies and compact stellar 
   systems (see Section \ref{Sec:ZoA}). The selection region extends to 1.5 mag beyond the edge of 
   the ``zone of avoidance" to ensure that we do not reject genuinely compact stellar systems. The large blue 
   stars are those objects which meet the selection criteria (including the ellipticity limit and a visual check 
   for obvious artefacts/background galaxies) and are therefore suitable for spectroscopic follow-up.}
   \label{fig:Selection}
\end{figure}

In this project we select CSS candidates using the following procedure:

\begin{enumerate}

\item We first search the Hyperleda catalog \citep{Hyperleda}\footnote{http://leda.univ-lyon1.fr/} 
to find all galaxies with recessional velocity between 500 and 14,000 km\,s$^{-1}$ ($\sim$ 7 - 200~Mpc 
assuming H$_{\rm 0}$=70 km\,s$^{-1}$Mpc$^{-1}$) with M$_{\rm B}$ $<$ --15. Given the resolution of 
the HST, this distance limit ensures that CSSs of effective radius $>$50~pc will be adequately resolved 
in any available HST images.

\item We then search the HST archive\footnote{http://archive.stsci.edu/hst/} for suitable WFPC2, 
ACS, and WFC3 broad band optical images (i.e. exposures in the W or LP filters with exposure 
times $>$ 50s, and at least two subexposures to allow adequate cosmic ray removal) located within 
150~kpc in projection of all the selected galaxies (only 9 out of 813 objects from the extended 
cluster/UCD catalog of \citealt{Bruens12} are located beyond 150~kpc from their host galaxy). This 
limitation is simply to ensure that we can safely make the necessary assumption that CSSs and the 
host galaxy (of known distance) are physically associated. This necessarily means that isolated 
CSSs like the one discovered by \citet{Huxor13} are unlikely to be discovered by our approach.

\item Suitable drizzled frames are then downloaded from the HST archive and analysed using 
SExtractor \citep{Bertin_sextractor}. SExtractor is used to produce a list of detected objects, their 
apparent magnitudes, a first estimate of their effective radii, and CLASS$\_$STAR (star-galaxy 
separation parameter) values. The principle benefits of using SExtractor to examine the images are 
the speed of the process and the reliability of the star-galaxy classification, which allows reliable 
rejection of unresolved objects without the need to individually fit surface brightness profiles for 
each object in the image. The principle limitation of using SExtractor is in the way backgrounds 
are subtracted: if the BACK$\_$SIZE parameter is too large CSSs located close to bright galaxies 
are lost, where BACK$\_$SIZE is set too small then the CSSs are themselves over subtracted leading 
to systematically reduced effective radii estimates. We have optimised the choice of BACK$\_$SIZE
using HST imaging of known CSSs as a training set, finding that a BACK$\_$SIZE of 64 is optimal
for ACS images and 32 for WFPC2.

\item The SExtractor catalogues produced from different pointings and instruments are then combined,
with overlapping magnitude estimates (e.g. from overlapping pointings with WFPC2 and ACS)
averaged with error weighting. For any particular filter where overlap occurs between instruments, 
ACS size estimates are preferred to WFPC2, and WFC3 is preferred to both ACS and WFPC2.

\item The photometry of every detected object is corrected for Galactic extinction (following \citealt{Schlafly11}) 
and then converted into an absolute magnitude assuming every object is located at the distance of 
the main galaxy. Likewise the size estimates are converted into physical units. 

\item A version of Figure \ref{fig:Selection} is produced for each filter available. Using the properties 
of known CSSs (from our master catalog described in Section \ref{Sec:LitComSam}) as a training set, 
a conservative region containing the rough mass--size trend of CSSs is then selected. This selection region 
extends 1.5 magnitudes beyond the approximate edge of the previously known CSS population to ensure 
that we do not reject genuinely more compact objects. Objects which lie within this region, are relatively 
round ($\epsilon$ $<$ 0.25), and have spatially resolved effective radii, are retained for further study. If 
subsequently spectroscopically confirmed, these first-pass estimates of the physical properties of the 
objects are refined using more sophisticated methods (see Section \ref{sec:halflight}).

\item We apply no colour cuts, thereby avoiding discarding potentially interesting objects that could
be young and blue, or dusty and red, such as the young massive clusters (YMCs) of NGC~34 and NGC~7252.

\item The remaining candidates are then examined individually by eye and any remaining spurious 
(due to artefacts, obvious background galaxies, etc.) candidates removed. Candidates are then 
cross-matched with literature compilations to prevent re-observation of previously known targets.
	
\end{enumerate}

Although we make every effort to be as complete as possible, there are particular situations where our 
photometric completeness is likely to be severely reduced. A first obvious case is for objects projected
close to the central regions of bright galaxies, where the high (and quickly varying) galaxy background
is difficult to subtract cleanly in an automated manner. A more subtle example is for objects associated 
with spiral galaxies. In this case the irregular galaxy light distribution makes reliably detecting CSSs
projected onto the face of the disk extremely challenging. Only in cases of edge-on spiral galaxies do 
we expect to reliably detect CSSs associated with spirals. It is also the case that our selection region 
could also potentially reject the youngest and most luminous YMCs, which if they are sufficiently massive
and young (but not enshrouded by dust) could lie more than 1.5 magnitudes into the ``zone of avoidance"
(see Section \ref{Sec:ZoA}) for old stellar systems. Finally in cases of galaxy mergers, the complex light 
distributions mean that there will be significant spatial variations of CSS detection efficiency.

The CSS candidates were then targeted for spectroscopic confirmation, principally at the SOAR and Keck 
telescopes. As these spectra were generally obtained during twilight or as filler targets, the objects for 
which spectra were obtained were generally brighter targets (V = 21.5 is the practical limit) selected randomly 
based on the current airmass (Table 1).

\begin{figure} 
   \centering
   \begin{turn}{0}
   \includegraphics[scale=0.9]{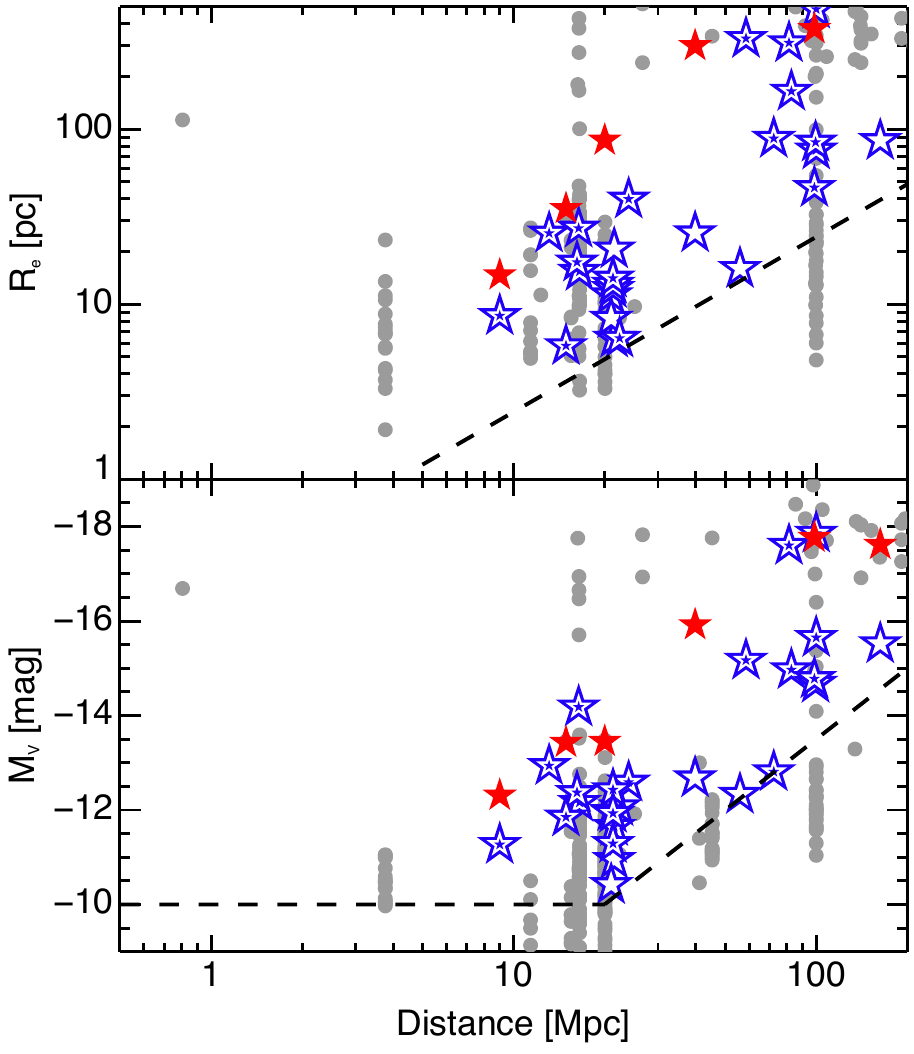}
   \end{turn} 
   \caption{$\bf{Upper~Panel:}$ The effect of our requirement that objects must be resolved by the HST
   before we conduct spectroscopic follow up. The blue stars are our confirmed objects, the ones with 
   filled stars inside denote the objects for which we were also able to obtain velocity dispersions. The 
   red stars denote objects previously known which were observed as a comparison sample, all of which
   meet the same selections as the main AIMSS sample. The grey circles are literature GCs, UCDs and
   cEs. The dashed line shows the resolution limit of the HST with distance, assuming conservatively that 
   to resolve an object, 2$\times$ the effective radius must be larger than the HST resolution limit of 
   0.1\arcsec. $\bf{Lower~Panel:}$ The effect of our requirement that objects for spectroscopic follow-up 
   must have M$_{\rm V}$ $<$ -10 and apparent V magnitude brighter than 21.5. The combination of the 
   two requirements leads to the dashed line.
   }
   \label{fig:Selection2}
\end{figure}

The net effect of the various selections, both in photometrically selecting targets, and in ensuring they 
were sufficiently luminous to spectroscopically confirm, is shown in Figure \ref{fig:Selection2}. The two
major limitations and their implications are; 1) the necessity of resolving the objects in the HST imaging, 
meaning that the allowed effective radius increases with redshift, and 2) the V=21.5 magnitude limit 
for spectroscopic followup, which means that only progressively brighter objects are found at higher redshift.

While conducting this project, several of our AIMSS targets were independently discovered and 
described by other authors (e.g. NGC~1132-UCD1: \citealt{Madrid11,Madrid13}, Perseus-UCD13: 
\citealt{Penny12}, M60-UCD1: \citealt{Strader13}). In what follows we make no distinction between 
these objects and the other AIMSS objects, as they were all selected using the above criteria and were 
analysed using the same methods.

\subsection{Literature Comparison Samples}
\label{Sec:LitComSam}

In addition to the AIMSS selected sample we include several complementary literature samples
that allow us to explore the properties of our objects relative to other compact stellar systems and
galaxy types. 

It is our intention to provide the most comprehensive catalog of CSS properties available. Therefore, 
we have attempted to include $\emph{all}$ spectroscopically confirmed UCDs and cEs in the literature which 
have available size measurements and which are more luminous than M$_{\rm V}$ = --10. The principle 
literature sources for CSSs are the compilations of \citet{Misgeld&Hilker11} and \citet{Brodie11} plus 
the recent update from \cite{Forbes13}, and to these compilations we add additional data for specific 
systems. Where possible we compile literature photometry for all objects and recompute their stellar 
masses using the same procedure as for our CSS sample (see Section \ref{Sec:StellMass}). We note 
below those cases where this is not possible due to limited available photometry. The
use of literature stellar masses ``as is'' can obviously lead to systematic offsets in the stellar masses
of some samples and/or object types. However, the magnitude of such offsets, for example by using
stellar masses derived using Kroupa instead of Salpeter IMFs, is around 40$\%$ which is small compared
to the factor of 2-3 errors introduced by for example assuming a single old SSP vs a composite age stellar 
population. Therefore we do not attempt to homogenise literature stellar mass estimates, especially as
this process itself could lead to additional errors, in particular due to the observed 
variation of IMF with stellar mass \cite[see e.g.][]{ATLAS3DXX}.

To our knowledge, our compilation of 191 objects is the most comprehensive catalog of GCs, UCDs, and cEs assembled
to date for objects that have been spectroscopically confirmed, are more luminous than M$_{\rm V}$ = -10, 
and have measured effective radii. The complete catalog of CSSs and comparison samples are
provided in Appendix \ref{Sec:Appendix}.

Throughout this paper we will discuss the properties of the CSS sample in relation to the early-type 
galaxies in our sample. However, we fully expect that later-type galaxies play an equally important 
role (possibly a dominant role in field/group environments) in the formation of certain types of CSS. We
defer a detailed discussion of late-type galaxies in order to simplify the analysis, and in the belief that 
the behaviour of partially rotationally supported systems such as dEs/dS0s and S0s likely encapsulates 
much of the behaviour of later type systems without the added complications to analysis caused by 
ongoing star formation and internal dust.

\subsubsection{dSph, dE and dS0 galaxies}

We include literature data for a sample of dwarf spheroidals, dwarf ellipticals, and dwarf S0s to 
allow their comparison with our UCD and cE samples. Data on the Milky Way and M31 dSph/dE 
systems come from \citet{Walker09}, \cite{McConnachie12}, \citet{Tollerud12}, \cite{Tollerud13}, 
and references therein. Data on the dE/dS0 sample comes from \citet{Geha02, Geha03}, 
\citet{Chilingarian09b}, and \cite{Toloba12} combined with five lower luminosity dwarf galaxies from 
\cite{Forbes11}. Where possible we add additional photometry for each source from the Sloan Digital 
Sky Survey DR9 \citep{SDSSDR9} corrected for foreground extinction following \cite{Schlafly11}, 
in order to improve the subsequent determination of the stellar mass (see Section \ref{Sec:StellMass}).

\subsubsection{Nuclear Star Clusters}

We make use of the compilations from \citet{Misgeld&Hilker11} and \citet{Brodie11}
to provide a comparison sample of dwarf galaxy nuclear star clusters.

\subsubsection{Massive Early-Type Galaxies}

In order to examine the connection between UCDs/cEs and early-type galaxies (both ellipticals and S0s) 
we make use of the galaxy sample from \citet{Misgeld&Hilker11} and \cite{Brodie11} combined with the 
ATLAS$^{\rm 3D}$ \citep{ATLAS3DI} survey to provide a representative comparison sample of normal 
early-type galaxies. Where galaxies are found in either \citet{Misgeld&Hilker11} or \citet{Brodie11} and also
in the ATLAS$^{\rm 3D}$ sample we prefer the ATLAS$^{\rm 3D}$ derived properties in our compilation.

We also take the effective radii and the total K-band magnitudes from \citet[][originally from 2MASS]{ATLAS3DI}. 
We then convert the K band integrated magnitudes to V assuming ${\rm{V-K} = 2.91}$, a colour that is 
appropriate for a stellar population with age = 10 Gyr and solar metallicity \citep{Maraston05}, and which should 
roughly match the stellar populations of the average elliptical galaxy. We derive stellar masses using the 
published stellar M/L$_{\rm r}$ values and r band luminosity from \citet{ATLAS3DXX}, where the M/L$_{\rm stars}$ taken from \citet{ATLAS3DXX} allows 
for some variation in the IMF between Kroupa and Salpeter (again, only a ~40$\%$ effect). The quoted 
central (R$_{\rm e}$/8) velocity dispersions are from \citet{ATLAS3DXV}.

\subsubsection{Young Massive Clusters}

In order to demonstrate the effects of age on the observed properties of compact stellar systems,
we include several young massive star clusters (YMCs) found in recent galaxy mergers. Specifically, we 
include W3, W6, W26, and W30 from NGC~7252; S$\&$S1 and S$\&$S2 from NGC~34; and 
G114 from NGC~1316. These young objects are expected to fade over several Gyrs to resemble 
UCDs \citep{Maraston04}.

The data for the YMCs comes mostly from the literature, although we spectroscopically reobserved 
W3 and W6 with SOAR as part of our calibration sample. The photometry and size estimates for the 
NGC~7252 clusters are from \citet{Bastian13}, those for the NGC~34 clusters are from \citet{Schweizer07},
and the measurement for NGC~1316-G114 is from \citet{Bastian06}. In addition to the size estimates
there are literature measurements for the velocity dispersions for NGC~7252 W3, W30, and NGC~1316 G114,
which come from \citet{Maraston04} and \citet{Bastian06} respectively.

\subsubsection{Globular Clusters}

We include the M$_{\rm V}$ and R$_{\rm e}$ values for Milky Way GCs from \cite{Brodie11} and add the 
measured central velocity dispersions of the clusters from \citet{Harris96} (2010 edition). We add the 
``extended but faint" GCs from \citet{Forbes13} and the M31 GCs from \cite{Strader11a}, where we convert 
their measured M$_{\rm K}$ to stellar mass assuming M/L$_{\rm K}$ = 0.937, which is appropriate for a 
stellar population with age = 10 Gyr and [Fe/H] $\sim$ --0.8, when assuming a Kroupa IMF
which better fits GCs than a Salpeter IMF \citep{Strader11a}. The effect of changing [Fe/H] by $\pm$ 0.5 
only results in M$_{\rm K}$ changing by 0.01 at this age \citep{Maraston05}. However, \citet{Strader09} 
and \citet{Strader11a} observe that those M31 GCs with [Fe/H] $<$ $-1$ have $M/L$ relatively 
consistent with a Kroupa IMF, whereas nearly all M31 GCs with [Fe/H] $>$ $-1$ have lower $M/L$ than predicted 
from stellar population models with a Kroupa IMF. The most likely reason is that these GCs have a 
deficiency of low-mass stars with respect to the assumed IMF, although standard dynamical evolution 
\citep{Kruijssen09} does not appear to  be a viable explanation for these observations. For our purpose, 
the main implication of assuming a fixed M/L$_{\rm K}$ is that the stellar masses of some of the GCs could
be in error by factors of 2--3. However, this does not affect any of the conclusions of the paper.

\subsubsection{UCDs and cEs}

We include additional UCDs from \cite{Norris&Kannappan11}, \citet{Norris12}, and \citet{Mieske13}, as well 
as additional cEs from \citet{Chilingarian10a} and \cite{Huxor11b,Huxor13}. Where possible we take central 
velocity dispersions rather than aperture measurements.

\begin{table*}
\begin{center}
\begin{tabular}{lllllll} \hline
Name				&	R.A		& Dec.		&Date		&Telescope	& Setup										& 	V$_{\rm helio}$	 	\\
					&	(J2000)	& (J2000)		&(dd/mm/yy)	&			&											&	(km/s)			\\
\hline
NGC 0524-AIMSS1		& 01:24:45.6	& +09:33:26.1	& 14/08/10	& SOAR		& GS 1200 1.68\arcsec\, 4800s 3.04\AA\, 0.9\arcsec		&	2446$\pm$18		\\
					& 			& 			& 09/11/10	& SOAR 		& GS  2100 0.84\arcsec\, 8400s 0.92\AA\, 0.7\arcsec	&	2525$\pm$8		\\					
NGC 0703-AIMSS1		& 01:52:41.1	& +36:10:14.4	& 20/10/12	& Keck		& DM 1200 1.00\arcsec\, 716s	1.55\,\AA\, 0.8\arcsec	&	5685$\pm$13		\\
NGC 0741-AIMSS1		& 01:56:21.3	& +05:37:46.8	& 12/01/13	& Keck		& DM 1200 1.00\arcsec\, 960s 1.55\,\AA\, 1.1\arcsec		&	5243$\pm$14		\\
NGC 0821-AIMSS1		& 02:08:20.7	& +10:59:26.6	& 23/10/06	& Keck		& DM 1200 1.00\arcsec\, 3600s 1.55\,\AA\, 1.0\arcsec	&	1705$\pm$6		\\
NGC 0821-AIMSS2		& 02:08:20.7	& +10:58:55.5	& 13/01/10	& Keck		& DM 1200 1.00\arcsec\, 5400s 1.55\,\AA\, 1.1\arcsec	&	1480$\pm$5		\\
NGC 0839-AIMSS1		& 02:09:40.6	&--\,10:11:07.1	& 07/11/12	& Keck		& ESI 0.5\arcsec\, 1200s 0.59\AA\, 0.7\arcsec			&	3791$\pm$34		\\
NGC 1128-AIMSS1		& 02:57:41.7	& +06:02:19.1 	& 13/01/13	& Keck		& DM 1200 1.00\arcsec\, 1200s 1.55\,\AA\, 1.2\arcsec	& 	7320$\pm$21		\\
NGC 1128-AIMSS2		& 02:57:44.5	& +06:02:02.2	& 13/01/13	& Keck		& DM 1200 1.00\arcsec\, 1200s 1.55\,\AA\, 1.2\arcsec	& 	7790$\pm$13		\\
NGC 1132-UCD1		& 02:52:51.2	& --\,01:16:18.8	& 28/10/11	& SOAR		& GS 2100 1.00\arcsec\, 3600s 1.08\AA\, 1.0\arcsec		&	7159$\pm$27		\\
NGC 1172-AIMSS1		& 03:01:36.4 	& --\,14:50:51.6	& 21/02/12	& Keck		& DM 900  1.00\arcsec\, 600s	2.2\,\AA\, 1.0\arcsec		&	1743$\pm$6		\\
NGC 1172-AIMSS2		& 03:01:34.4	& --\,14:49:50.7	& 21/02/12	& Keck		& DM 900  1.00\arcsec\, 600s	2.2\,\AA\, 1.0\arcsec		&	1617$\pm$15		\\
Perseus-UCD13		& 03:19:45.1	& +41:32:06.0	& 20/02/12	& Keck		& DM 900 1.00\arcsec\, 4800s	2.2\,\AA\, 1.0\arcsec		&	5292$\pm$14	 	\\	
NGC 1316-AIMSS1		& 03:22:36.5	& --\,37:10:55.9 & 26/10/11	& SOAR		& GS 2100 1.00\arcsec\, 4800s 1.05\AA\, 0.8\arcsec		&	1976$\pm$12		\\
NGC 1316-AIMSS2		& 03:22:33.3	& --\,37:11:13.1 & 26/10/11	& SOAR		& GS 2100 1.00\arcsec\, 4800s 1.05\AA\, 0.8\arcsec		&	1396$\pm$13		\\
NGC 2768-AIMSS1		& 09:11:36.8	& +60:04:16.1	& 08/11/12	& Keck		& ESI 0.5\arcsec\, 900s	 0.59\AA\, 0.8\arcsec			&	 1214$\pm$20		\\
NGC 2832-AIMSS1		& 09:19:46.3	& +33:45:46.5	& 04/12/11	& Keck		& DM 1200 1.00\arcsec\, 1.55\AA\, 1.0\arcsec			&	6607$\pm$20		\\	
NGC 3115-AIMSS1		& 10:05:15.8	& --\,07:42:51.6	& 07/11/12	& Keck		& ESI 0.5\arcsec\, 1800s 0.59\AA\, 1.2\arcsec			&	 726$\pm$19		\\
NGC 3268-AIMSS1		& 10:30:00.1	& --\,35:20:19.4	& 27-8/01/12	& SALT		& RSS 2300 2.00\arcsec\,	 3000s  2.16\AA\, 1.5\arcsec	&	2455$\pm$57		\\		
NGC 3923-UCD1		& 11:51:04.1	& --\,28:48:19.8	& 15/04/09	& SOAR		& GS 600 1.5\arcsec\, 9600s 6.2\AA\, 0.6\arcsec		&	2097$\pm$18$^\dagger$		\\
					& 			& 			& 30/04/11	& Gemini-S	& GM-S 1200 0.5\arcsec\, 10688s 1.26\AA\, 0.9\arcsec	&	2115$\pm$30$^\diamond$ 	\\
NGC 3923-UCD2		& 11:50:55.9	& --\,28:48:18.4	& 15/04/09	& SOAR		& GS 600 1.5\arcsec\, 9600s 6.2\AA\, 0.6\arcsec		&	1501$\pm$44$^\dagger$		\\
					& 			& 			& 30/04/11	& Gemini-S	& GM-S 1200 0.5\arcsec\, 10688s 1.25\AA\, 0.9\arcsec	&	1478$\pm$29$^\diamond$ 	\\
NGC 3923-UCD3		& 11:51:05.2	& --\,28:48:58.9	& 30/04/11	& Gemini-S	& GM-S 1200 0.5\arcsec\, 10688s 1.26\AA\, 0.9\arcsec	&	2308$\pm$35$^\diamond$ 	\\
NGC 4350-AIMSS1		& 12:23:59.1	& +16:41:07.9 	& 20/02/12	& Keck		& DM 1200 1.0\arcsec\, 900s 1.55\,\AA\, 0.9\arcsec		&	1183$\pm$8		\\
NGC 4546-UCD1		& 12:35:28.7	& --\,03:47:21.1	& 18/04/09	& SOAR		& GS 600 1.68\arcsec\, 7200s 6.3\AA\, 0.6\arcsec		&	1256$\pm$24$^\dagger$		\\
					& 			& 			& 04/03/11	& SOAR		& GS  2100 1.03\arcsec\, 3600s 1.13\AA\, 1.2\arcsec	&	1182$\pm$2 		\\
NGC 4565-AIMSS1		& 12:36:37.2	&  +25:57:44.3 	& 05/03/13	& Keck		& ESI 0.5\arcsec\, 600s 0.59\AA\, 1.0\arcsec			&	1335$\pm$9		\\
NGC 4621-AIMSS1		& 12:41:52.9 	& +11:37:47.9	& 11/01/13	& Keck		& DM 1200 1.00\arcsec\, 390s 1.55\,\AA\, 0.8\arcsec		&	474$\pm$6		\\
M60-UCD1			& 12:43:36.0 	& +11:32:04.6	& 11/01/12	& INT		& IDS R300V 1.00\arcsec\,1800s 4.12\,\AA\,			&	1236$\pm$33		\\	
					& 		 	& 			& 05/03/12	& SOAR		& GS 2100 1.00\arcsec\, 3600s 1.08\AA\, 1.0\arcsec		&	1258$\pm$11		\\	
ESO 383-G076-AIMSS1	& 13:47:25.4	& --\,32:52:56.3	& 29/01/12	& SALT		& RSS 2300 2.00\arcsec\,	1800s 2.16\AA\, 1.0\arcsec	&	11403$\pm$24		\\
NGC 7014-AIMSS1		& 21:07:51.5	& --\,47:11:25.6& 19/04/12		& SOAR		& GS 2100 1.00\arcsec\, 3600s 1.05\AA\, 0.7\arcsec		&	5197$\pm$14		\\
					&			& 			& 			& 			& 											&					\\

\multicolumn{6}{|l}{\bf Contaminants}&\\
NGC 7418A-BG1		& 22:56:43.2	& --\,36:46:43.1 & 02/05/11	& SOAR		& GS 2100 1.68\arcsec\, 3600s 1.19\AA			&	75300$\pm$60		\\

\hline
\end{tabular}
\caption[Redshifts]{Compact stellar systems spectroscopically confirmed by the AIMSS collaboration.  The setup
column describes the instrument (GS = Goodman Spectrograph, GM-S = GMOS South, DM = DEIMOS), grating (l/mm), slit width,
total exposure time and resulting FWHM spectral resolution in \AA, and seeing in arcseconds. \\
$^\dagger$ Previously reported in \citet{Norris&Kannappan11} \\
$^\diamond$ Previously reported in \citet{Norris12}.}
\end{center}
\label{tab:cztab}
\end{table*}

\begin{table*}
\begin{center}
\begin{tabular}{lllllll} \hline
Name				&	R.A		& Dec.		& Date		& Telescope	& Setup											& 	V$_{\rm helio}$	 	\\
					&	(J2000)	& (J2000)		& (dd/mm/yy)	&			&												&	(km/s)			\\
\hline
Fornax-UCD3			& 03:38:54.0	& --\,35:33:34.0	& 31/10/10	& SOAR		& GS 2100 1.5" 3600s 1.36\,\AA\, 1.0\arcsec				& 1473$\pm$6			\\ 
NGC 2832-cE			& 09:19:47.9	& +33:46:04.9	& 04/12/11	& Keck		& DEIMOS 1200 1.00\arcsec\, 1.55\,\AA\, 1.0\arcsec			& 7076$\pm$9			\\ 
NGC 2892-AIMSS1		& 09:32:53.9	& +67:36:54.5	&			& SDSS		& Spectrum from SDSS DR10							& 6847$\pm$3			\\
NGC 3268-cE1/FS90 192	& 10:30:05.1	& --\,35:20:32.0	& 27-8/01/12	& SALT		& RSS 2300 2.00\arcsec\,	 3000s 2.16\AA	\, 1.5\arcsec		& 2479$\pm$27		\\
Sombrero-UCD1		& 12:40:03.1	& --\,11:40:04.3	& 05/03/11	& SOAR		& GS 2100 1.03" 2700s 1.08\,\AA\, 1.1\arcsec				& 1306$\pm$6			\\
M59cO				& 12:41:55.3 	& +11:40:03.8	& 11/01/13	& Keck		& DEIMOS 1200 1.00\arcsec\, 390s 1.55\,\AA\, 1.0\arcsec		& 703$\pm$9			\\
ESO 383-G076-AIMSS2	& 13:47:25.3	& --\,32:53:09.9	& 29/01/12	& SALT		& RSS 2300 2.00\arcsec\, 1800s 2.16\AA\, 1.0\arcsec		& 10978$\pm$8		\\	
NGC 7252-W3			& 22:20:43.7 	& --\,24:40:38.0	& 30/05/11	& SOAR		& GS 2100 1.03" 3600s 1.05\,\AA\, 0.8\arcsec				& 4744$\pm$12		\\
NGC 7252-W6			& 22:20:44.0 	& --\,24:40:27.7	& 30/05/11	& SOAR		& GS 2100 1.03" 3600s 1.05\,\AA\,0.8\arcsec				& 4606$\pm$9			\\

\hline
\end{tabular}
\caption[Redshifts]{Previously confirmed compact stellar systems and serendipitous objects (those observed in the same slit as AIMSS objects) 
with additional observations/analysis provided by the AIMSS collaboration.}
\end{center}
\label{tab:cztab2}
\end{table*}

\section{Observational Methods}

\subsection{Spectroscopic Observations}
Tables 1 and 2 present the observing logs of objects observed to date as part of the AIMSS project. 
Table 1 shows those objects newly discovered or confirmed by this project, while Table 2 shows a 
sample of previously confirmed objects re-observed by us to provide a comparison/calibration sample. 
In both tables column 1 is a designation for the object, columns 2 $\&$ 3 are the Right Ascension 
and Declination in J2000 coordinates, column 3 is the date of observation, column 4 lists the 
Telescope used to determine the objects redshift, column 5 provides the instrument setup used 
(instrument, grating, slit width, exposure time, and spectral resolution FWHM), and column 6 gives 
the heliocentric recessional velocity and its error rounded to the nearest km\,s$^{-1}$ measured 
using the procedure described in Section \ref{sec:cz_determination}.

\subsubsection{SOAR}
The majority of our southern spectroscopic observations to date have been obtained using the 
Southern Astrophysical Research (SOAR) Telescope and the Goodman spectrograph 
\citep{GoodmanSpec} in longslit and MOS modes. Our preferred setup with a 1\arcsec\, 
longslit and the 2100l/mm VPH grating provides spectral resolution of FWHM $\sim$1\AA\, 
sampled at 0.33\AA\, with spectral coverage from 4850 to 5500\AA.

\subsubsection{SALT}
We used the South African Large Telescope (SALT) to observe fainter targets requiring exposure
times impractically long to be used as filler targets for SOAR observing and which cannot be 
observed with Keck. The observations used the RSS spectrograph \citep{RSS} with the 2300l/mm 
grating and a 2\arcsec\, wide longslit providing coverage from $\sim$4300 to 7400\AA\, with a 
spectral resolution (FWHM) of 2.16\AA, sampled with 0.7\AA\, pixels.

\subsubsection{Gemini-South}
As part of a study examining the GCs and UCDs of the shell elliptical NGC~3923 we obtained 
deep Gemini/GMOS spectroscopy of three UCDs (see \citealt{Norris12} for further details). The
observations were made in MOS mode with the 1200l/mm grating and 0.5\arcsec slitlets, yielding 
spectra with a resolution of 1.26\AA\, FWHM and wavelength coverage from $\sim$4100 to 5600\AA\,. 
These observations were sufficiently deep to allow the measurement of velocity dispersions for 
all three UCDs.

\subsubsection{Keck}
The majority of our northern hemisphere candidates were spectroscopically confirmed using 
the DEIMOS and ESI instruments on the Keck telescope \citep{DEIMOS,ESI}. For DEIMOS
our observational setup uses the 1200 grating, and a 1\arcsec\, wide longslit, centred on the 
Calcium triplet region ($\sim$8500\AA) providing a spectral resolution of 1.55\AA\, sampled 
at 0.32\AA. ESI gives a coverage from 3900 to 10900\AA\, and for our observations provides 
a spectral resolution of $\sim$0.59\AA\, when using a 0.5\arcsec\, wide longslit.

\subsubsection{INT}
We also obtained spectra of NGC~4649 UCD1 with the IDS instrument on the Isaac Newton
Telescope using the RED+2 detector, the R300V grating, and a 1\arcsec\, longslit, providing
a resolution of 4.12\AA\, FWHM over the whole visible spectrum.

\subsection{Spectral Reduction}

Where available, our spectroscopic observations were reduced using the dedicated pipelines 
of the particular instruments used, e.g. using the Gemini-GMOS \textsc{IRAF} package for GMOS 
observations as described in \cite{Norris12}.

In those cases where no dedicated pipeline currently exists, or it is insufficient for our purposes 
(i.e. for SOAR-Goodman, INT, and SALT-RSS spectroscopy), the observations were reduced using 
custom reduction pipelines. The pipelines used standard \textsc{IRAF} routines to carry out bias 
and overscan subtraction, trimming of the science data to remove unnecessary spatial coverage, 
then flat fielding. Following flat fielding the \textsc{IDL} implementation of \textsc{la$\_$cosmic} 
\citep{lacosmic} was used to clean cosmic rays from each science exposure. \textsc{IRAF} was 
then used to carry out wavelength calibration and rectification, as well as object tracing and 
extraction into a 1D spectrum (using \textsc{apall}) and finally combination of individual exposures 
(\textsc{scombine}).

\subsection{Redshift Determination and Candidate Confirmation}
\label{sec:cz_determination}

We measure the redshifts of all of our CSSs by cross correlating the input spectra against 
a simple stellar population spectral library (the high resolution, FWHM = 0.55\AA, ELODIE based models of 
\citealt{MarastonStromback11}) in the case of optical spectra, and a library of empirical stellar spectra 
observed with the same setup used for the Keck/DEIMOS CaT observations. The cross correlation 
is carried out using the \textsc{IRAF} task \textsc{fxcor} in the \textsc{rv} package. More details 
regarding the estimation of errors and the procedure used to reject outlying velocities can be 
found in \cite{Norris&Kannappan11}. 

\begin{figure} 
   \centering
   \begin{turn}{0}
   \includegraphics[scale=1.0]{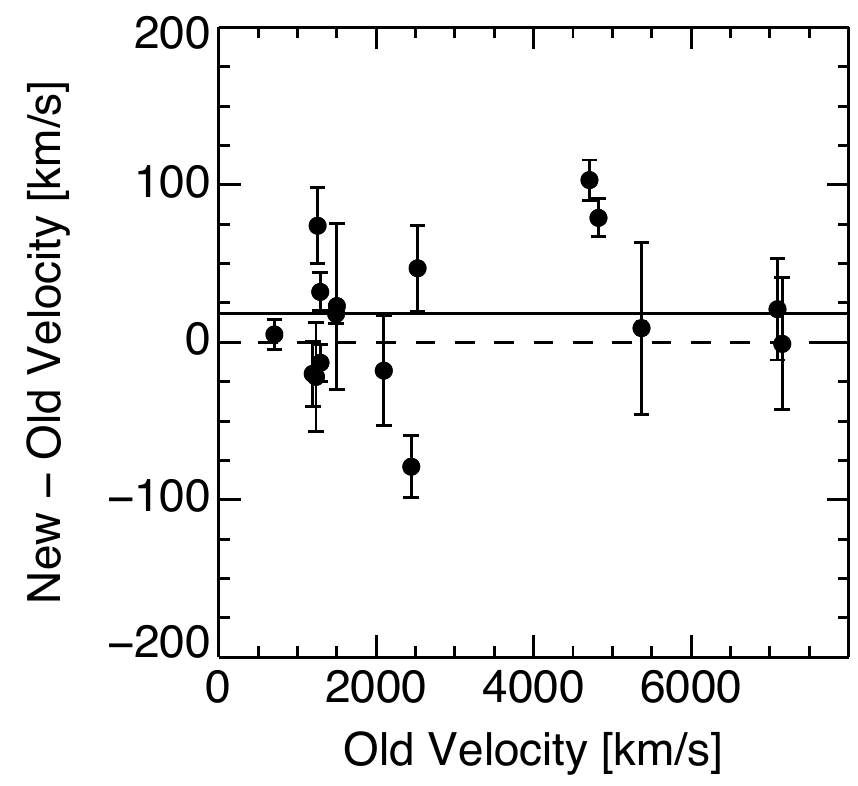}
   \end{turn} 
   \caption{Our repeat/new recessional velocities compared to earlier AIMSS or literature velocities for
   a sample of 15 objects with repeat spectroscopic observations. The dashed line is the 
   equality line, while the solid line shows the median of old-new velocities. The median 
   offset is 18 km/s, showing that our inhomogeneous spectroscopic observations are not
   systematically offset from previous measurements.
    }
   \label{fig:Velocities}
\end{figure}

Because we have re-observed several of our objects with various telescope and instrument 
combinations, as well as re-observing several calibration objects, we are able to determine the 
repeatability of our recessional velocity determinations. Figure \ref{fig:Velocities} demonstrates that our velocity 
repeatability is generally very good, across the wide variety of telescope and instrument configurations
which could lead to systematic differences between observations. The median difference between 
repeat measurements is 18 km\,s$^{\rm -1}$ with a dispersion of 45 km\,s$^{\rm -1}$ and all but two 
observations agree to within 3$\sigma$ of their respective errors. The two significant outliers are the 
young massive clusters of NGC~7252 (W3 and W6) which are expected to be problematic due to their 
very young ages \citep[$\sim$300Myr; ][]{Maraston04} which can lead to significant template mismatch. 
We therefore believe that our recessional velocities are reliable at the $\sim$50 km\,s$^{\rm -1}$ level, 
which is sufficient to determine physical association between CSS candidate and host, but possibly 
not accurate enough for detailed analyses of correlated structures in position-velocity phase space
(see e.g. \citealt{Romanowsky12}).

\begin{figure} 
   \centering
   \begin{turn}{0}
   \includegraphics[scale=1.0]{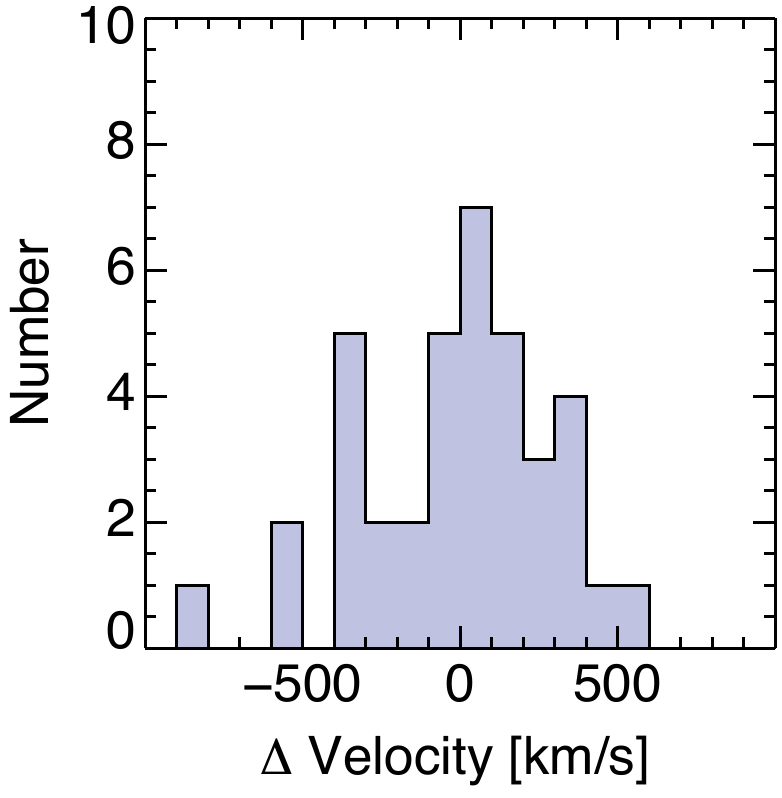}
   \end{turn} 
   \caption{Histogram of the $\Delta$ velocities (CSS recessional velocity - host galaxy recessional 
   velocity). The maximum velocity difference between CSS and presumed host is 849 km\,s$^{-1}$ for
   the cE of NGC~1128, which resides in a medium sized group. From this plot it can be seen that only
   the cE of NGC~1128 has a velocity offset larger than the largest velocity outlier in the GC system 
   of the Sombrero galaxy (550 km/s).}
   \label{fig:Redshift_Limits}
\end{figure}

To confirm the nature of candidate objects we examine the difference in redshift between the
CSS and the mean recessional velocity of the presumed host structure (i.e. galaxy, group, or 
cluster). Figure \ref{fig:Redshift_Limits} shows that only one AIMSS candidate found so far has 
a velocity offset greater than 550 km\,s$^{\rm -1}$ (except for one obvious high-z background object), 
which is similar to the largest velocity offset found for GCs in the GC system of the Sombrero 
galaxy \citep{Bridges07}, and slightly lower than the 650 km\,s$^{\rm -1}$ maximum offset found 
for GCs of the group elliptical NGC~3923 \citep{Norris08,Norris12}. 

\begin{figure} 
   \centering
   \begin{turn}{0}
   \includegraphics[scale=1.0]{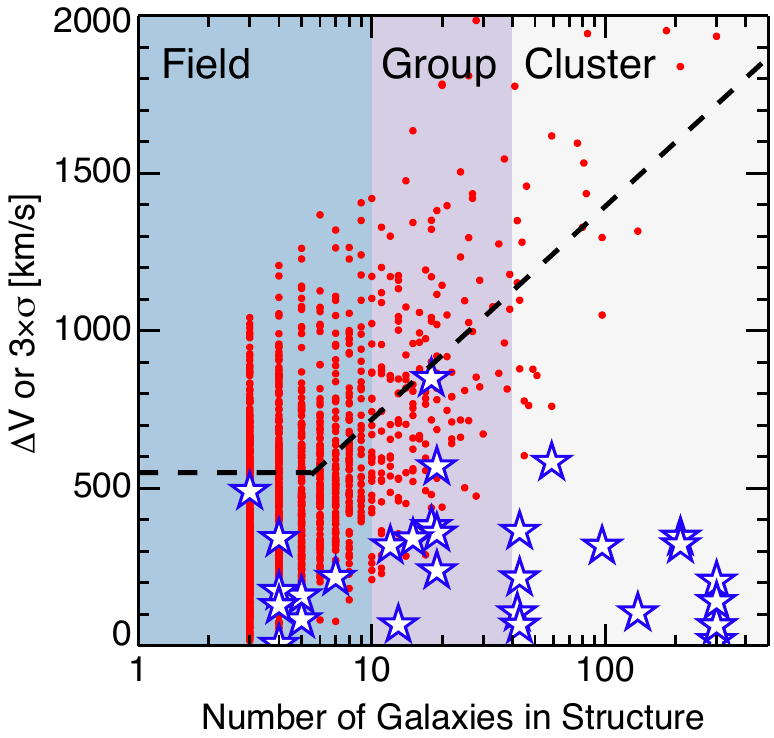}
  \end{turn} 
   \caption{Absolute velocity offset of CSS from host galaxy, or 3$\times$ the velocity dispersion of the 
   group versus the number of galaxies in the group from the low density contrast catalog of \citet{Crook07}. 
   The shaded regions display our adopted field/group/cluster classification. The blue stars are the velocity 
   offset between our confirmed AIMSS objects and their host galaxies, in the case where their host galaxy 
   is found in the \citet{Crook07} catalog. The red dots show 3$\times$ the global (group/cluster) velocity 
   dispersion of all groups found in the \citet{Crook07} catalog. 
   They can be thought of as the largest velocity offset from the structure mean (a 3 $\sigma$ outlier) 
   likely to be found for a galaxy within the bound structure. Hence CSS's selected to lie within this limit are likely 
   to be bound to the structure they are projected onto.
   The dashed line is a fit to the red points for groups with more than 5 
   members; below this it is a fixed value of 550 km/s chosen to match the largest expected velocity outlier 
   in the GC system of isolated mid-sized galaxies, such as the Sombrero galaxy. To date only one candidate 
   (an obvious background galaxy with cz $\sim$ 75000 km\,s$^{-1}$) has failed to lie below the dashed line 
   and therefore to be physically associated with the assumed host galaxy. There is a noticeable absence 
   of objects with velocity offsets above 400 km\,s$^{-1}$ for larger structures (Number of Galaxies $>$ 100).
   This may be an indication of the formation process; star cluster type objects will be expected to have 
   velocities close to their host galaxies, but objects formed by stripping also must have velocities similar to 
   those of the larger galaxies that did the stripping, as multiple close passes are required to do the 
   necessary stripping.   }
   \label{fig:Redshift_Limits2}
\end{figure}

To produce a systematic recessional velocity offset limit we make use of the 2MASS All Sky Redshift 
Survey group catalog of \citet{Crook07} which is the most complete over the whole sky and which uses 
a luminosity function correction to account for galaxies which fall below the magnitude limit of
the input redshift catalog. Figure \ref{fig:Redshift_Limits2} shows how we derive this limit. 
We start by plotting 3$\times$ the velocity dispersion of the group/cluster vs the total number of 
galaxies in the structure for all 1604 groups in the \citet{Crook07} low contrast group catalog 
(red points). We then fit a relation to these groups for all structures with more than 5 members 
(to ensure a reliable dispersion measurement). For structures with less than 5 members we allow 
a maximum velocity offset of 550 km\,s$^{\rm -1}$ as found for the Sombrero galaxy GC system, 
which produces the dashed lines in the plot. 

Where the host galaxy of the AIMSS candidate lacks a counterpart in the \citet{Crook07} catalog we 
make use of literature determinations of environment and assume that environments classified as 
``field" in the literature have 10 members, ``groups" have 40, and ``clusters" have 300. We then use
the derived maximum velocity offsets for structures of these sizes as the appropriate limits. We have 
overplotted the absolute velocity offset for all AIMSS objects which have host galaxies in the \citet{Crook07} 
catalog as blue stars. As can be seen all but one (NGC~7418A-BG1 not plotted) of our candidate 
CSSs are found to be physically associated with the assumed structure, leading to a total success 
rate of 96$\%$.

\subsection{Velocity Dispersion Determination}
\label{sec:sigma_determination}

Integrated velocity dispersions ($\sigma$) for our CSSs were measured where the available spectra had 
sufficient resolution and S/N (generally $>$ 25 per \AA\, was required to achieve reliable measurements), 
using version 4.65 of the penalised pixel fitting code (\textsc{ppxf}) of \cite{ppxf}. This code fits each input 
spectrum with an optimal combination of template (the same SSP models and stellar templates as described 
in Section \ref{sec:cz_determination}) spectra convolved with the line of sight velocity distribution (LOSVD). 

\begin{figure} 
   \centering
   \begin{turn}{0}
   \includegraphics[scale=1.00]{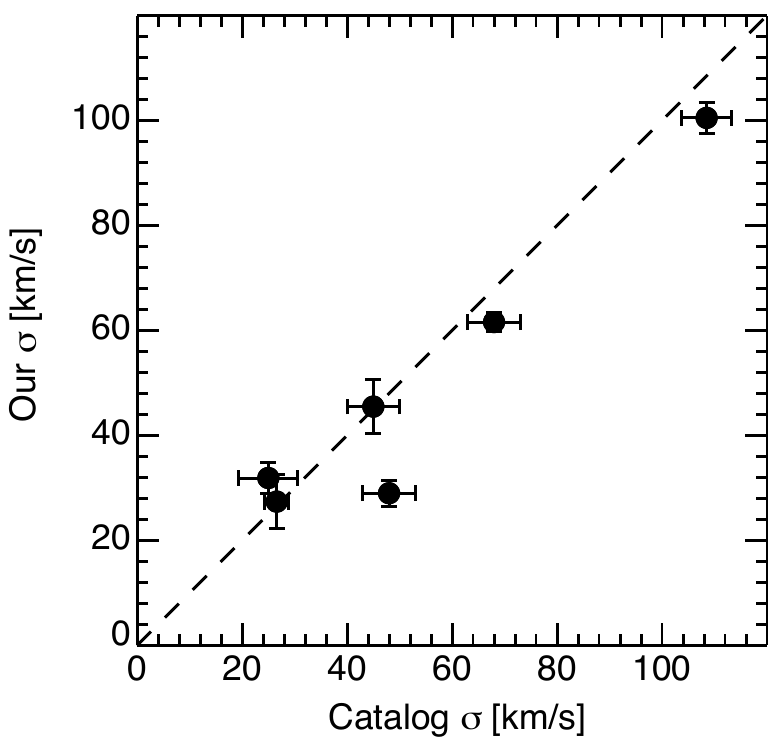}
   \end{turn} 
   \caption{Our AIMSS velocity dispersion measurements compared to literature values for 6 objects 
   which had previously been observed (from left-to-right; Sombrero-UCD1: \citealt{Hau09},
   Fornax-UCD3: \citealt{Mieske13}, NGC~7252-W3: \citealt{Maraston04}, M59cO: \citealt{Chilingarian08a},
   M60-UCD1: \citealt{Strader13}, NGC~2832-cE: \citealt{SDSSDR9}). The significant outlier is M59cO 
   (see Section \ref{sec:sigma_determination}). 
   The dashed line is the equality relation.  }
   \label{fig:Velocity_Dispersion}
\end{figure}

\begin{figure} 
   \centering
   \begin{turn}{0}
   \includegraphics[scale=0.9]{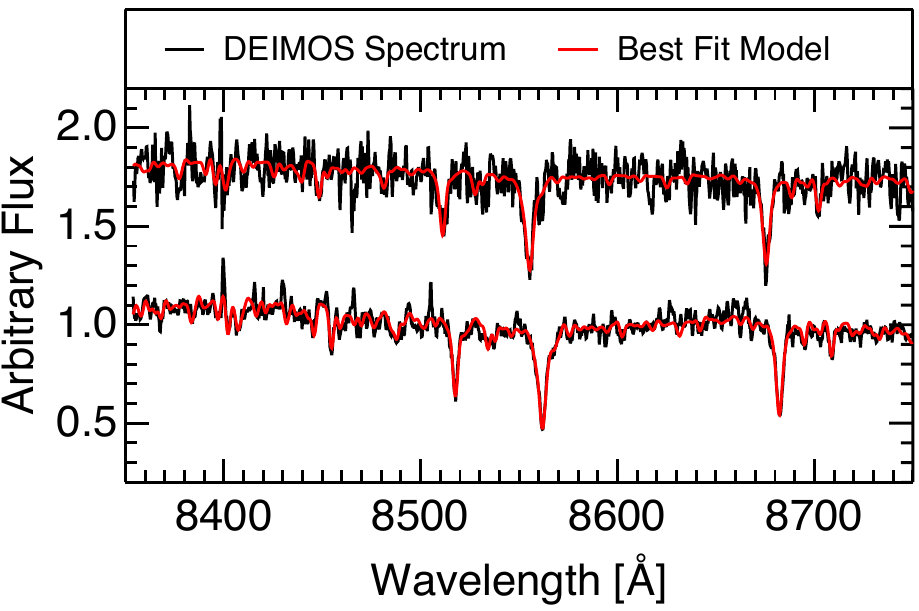}
   \end{turn} 
   \caption{Our Keck/DEIMOS spectra for NGC~4621-AIMSS1 (upper black spectrum) and M59cO (lower 
   black spectrum). The red lines in both cases are the best fit pPXF spectra. The actual flux values are 
   arbitrary, with the NGC~4621-AIMSS1 spectrum offset for clarity. The quality of the spectra and the pPXF 
   fits are evident in both cases.}
   \label{fig:M59cO_Spectrum}
\end{figure}

Figure \ref{fig:Velocity_Dispersion} shows our $\sigma$ measurements compared to literature 
measurements for 5 objects we re-observed as calibration objects. The one significant outlier is 
M59cO, where the literature value of 48 $\pm$ 5 km\,s$^{-1}$ from \citet[][]{Chilingarian08a} disagrees 
with ours (29.0 $\pm$ 2.5 km\,s$^{-1}$) by almost 3 standard deviations. As our pPXF fit to this spectrum 
is excellent (see Fig. \ref{fig:M59cO_Spectrum}), and the resolution of our spectrum (23 km\,s$^{-1}$ FWHM)
is significantly below the measured value, we choose to adopt our value. In this case the offset
is likely due to a combination of effects, including differences in seeing and slit/fibre width and positioning 
resulting in different spatial sampling and differences in the mix of templates used to fit
the spectrum. In addition, the \citet[][]{Chilingarian08a} value is derived from an SDSS 
spectrum and hence the resolution (of around 70 km\,s$^{-1}$) is significantly higher than the measured 
velocity dispersion. Thus, this measurement is likely more uncertain than the quoted error would imply. 
All other repeat measurements are within the mutual 1$\sigma$ errors, indicating that our measured 
velocity dispersions can be safely combined with other literature samples.

As mentioned above, a complication of the velocity dispersion determination is that we are sensitive 
to only the light which falls within the instrument longslit. This means that for strongly peaked velocity 
dispersion profiles such as those measured for UCDs and cEs \citep[e.g.][]{Chilingarian10a,Frank11}, 
the velocity dispersion we have determined is, in fact, a luminosity-weighted average between the central 
velocity dispersion, and the true global average velocity dispersion of the CSS. Therefore in order to
properly estimate the dynamical mass of our sample it is first necessary to model the intrinsic light
distribution of the CSS and then correct the measured velocity dispersion for the effects of slit losses 
and seeing. As the examination of the dynamical masses and mass-to-light ratios of our CSSs will
take place in a forthcoming paper (Forbes et al. 2014, in prep), we leave this additional analysis until 
then. For the current paper we treat our measured velocity dispersions as approximations to the central 
velocity dispersions, and expect that they are correct to within 10$\%$ of the final value \citep[see e.g.][]{Mieske08}.

\subsection{Photometric Reanalysis}

\subsubsection{Effective Radii}
\label{sec:halflight}

For those objects spectroscopically confirmed as compact stellar systems, we use the available
HST images to remeasure the effective radii using a range of techniques. We first subtract the 
host galaxy background. Where possible we use ELLIPSE in IRAF to model the galaxy background 
and remove it, after masking other objects within the HST field-of-view. In those cases where the 
host galaxy background cannot be adequately modelled using ELLIPSE (e.g. where the centre of 
the host galaxy is not located on the image) we produce a median smoothed image following the 
procedure outlined in \cite{Norris&Kannappan11}.

\begin{figure} 
   \centering
   \begin{turn}{0}
   \includegraphics[scale=1.0]{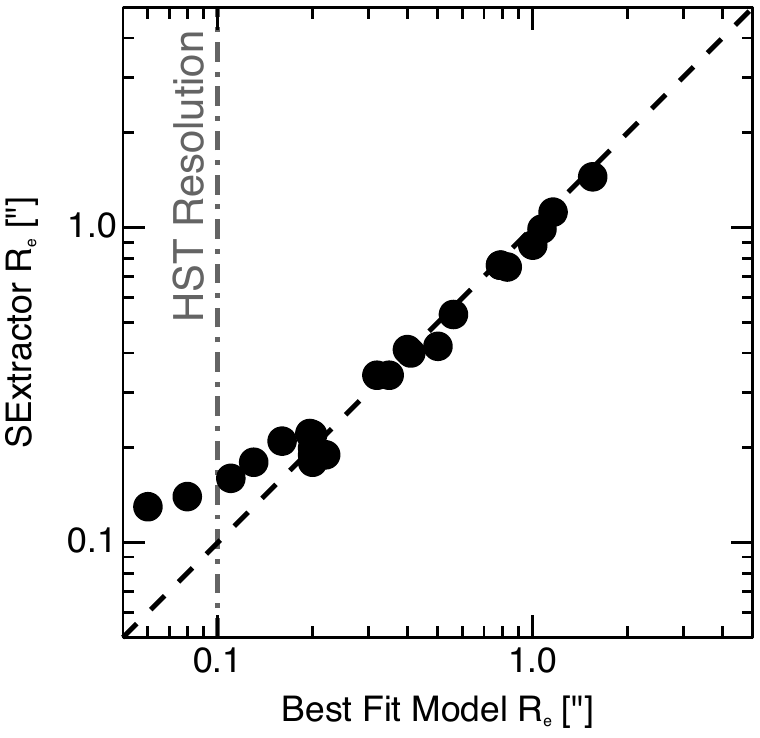}
   \end{turn} 
   \caption{Comparison of \Re\, determined using structural models fitted with ISHAPE (i.e. King, Sersic), 
   vs. those determined using SExtractor. The dashed black line is the one-to-one relation, the vertical 
   dot-dashed line shows the resolution limit of the HST WFPC2 and ACS cameras ($\sim$0.1$^{\prime\prime}$).
   It can be seen that for objects with \Re\, $>$ 3 HST resolution elements,
   SExtractor provides a reliable estimate of \Re.}
   \label{fig:Re_Comparison}
\end{figure}

After background subtraction we use SExtractor to produce a size estimate as a first guess input 
for the ISHAPE structural fitting code \citep{IShape99}. We then use ISHAPE to fit Sersic and King 
models (with concentration 15, 30, 100, and unconstrained) to each object, using a PSF constructed 
using TinyTim \citep{Krist11}. Where the best fit King or Sersic model (as judged by the ISHAPE 
$\chi^2$ value) has a radius less than 0.3\arcsec\, we accept this value as the correct major axis 
effective radius. For those cases where the best fit Sersic or King model has a radius greater than 
0.3\arcsec\, we use the SExtractor value, as this value is model independent and therefore potentially 
more resistant to under or over-fitting low surface brightness outer structures. 
Figure \ref{fig:Re_Comparison} demonstrates that in the case where R$_{\rm e}$ $>$ 0.3\arcsec\, 
(which is 3$\times$ the FWHM of the HST optical PSF) the ISHAPE and SExtractor estimates are 
in good agreement, with a median offset of $\sim$4$\%$. 

\begin{figure*} 
   \centering
   \begin{turn}{0}
   \includegraphics[scale=1.00]{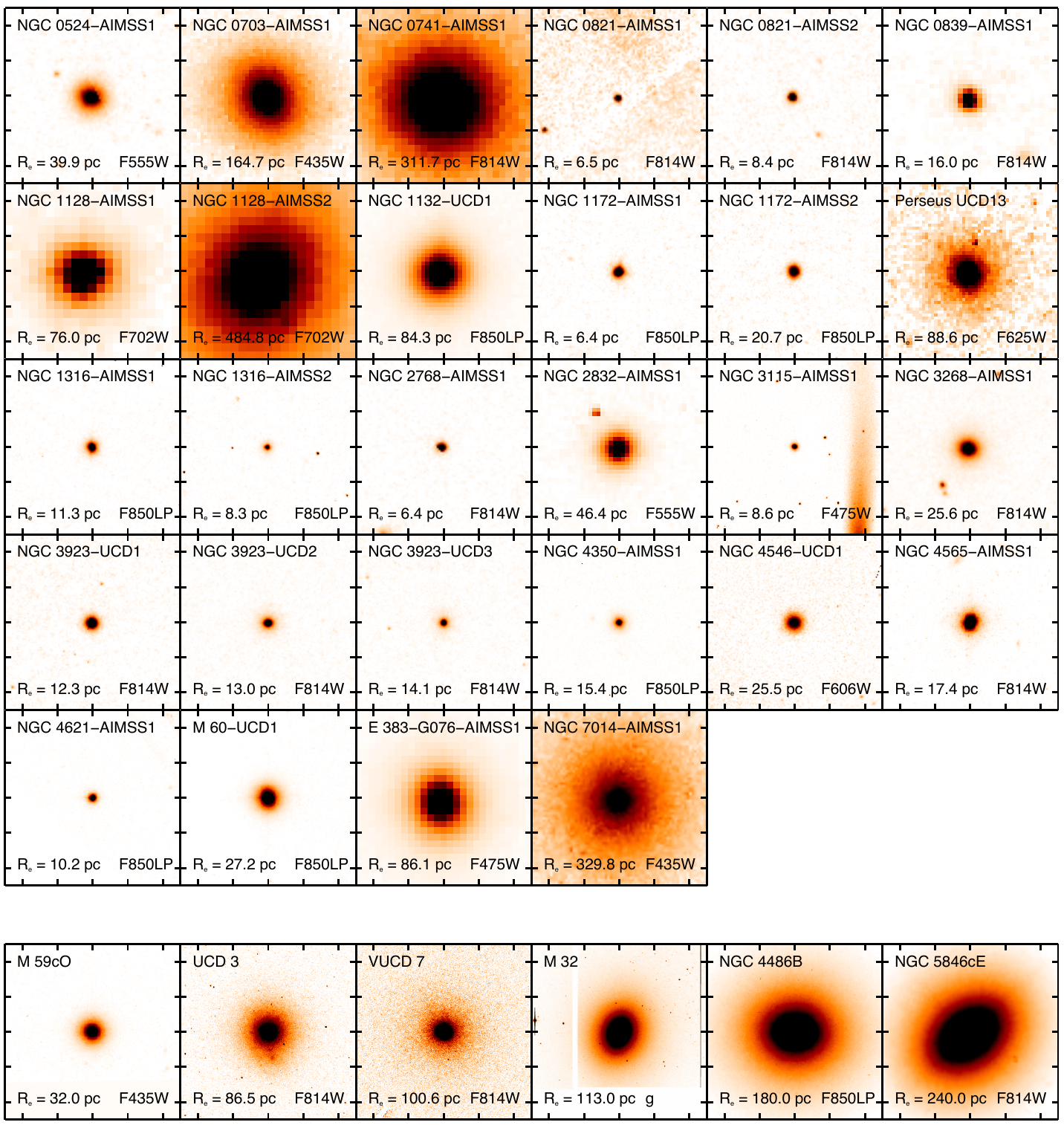}
   \end{turn} 
   \caption{Upper Panels: Thumbnails of our CSS sample. Each thumbnail is 1 $\times$ 1 kpc. The measured 
   effective radius for each CSS is provided in the bottom left of each panel, the filter of the image used to produce 
   the thumbnail is given in the bottom right. In all cases except M 32 the imaging is from the HST, for M 32 a 
   g band MegaPrime image is used due to the large size of M 32 on the sky.
   It is clear from these images that the more extended objects (those with R$_{\rm e}$ $\ga$ 30 pc) often appear to have 
   additional lower surface brightness outer components. Lower Panels: Six literature CSSs to provide a comparison
   sample.}
   \label{fig:Thumbnails}
\end{figure*}

Figure \ref{fig:Thumbnails} shows 1 $\times$ 1 kpc thumbnails for each of our sample. It is clear from
this figure the range of half-light radii displayed by our CSSs is significant. Also obvious is the fact that 
some of our larger (\Re\, $>$ 30 pc) objects show evidence for a multi-component structure, with signs 
of low surface brightness outer structures that maybe provide insights into their formation mechanisms.
The structural fitting analyses support this observation, with the larger objects often being poorly fit by
single component models, further validating our decision to use model independent effective radii where 
possible. For the present paper we leave off investigating the detailed structures of our objects, relying 
only on the simple (usually non-parametric) estimate of \Re\, described above which can most easily be 
compared to literature samples.

In order to allow consistent comparison of our data with the literature samples we have re-estimated the 
sizes of the seven compact Coma cluster objects from \citet{Price09}, because the \Re\, values given in 
\citet{Price09} are provided for two component structural models separately, and not for the total light 
distribution, and are therefore unsuitable for comparison with other literature data. Our remeasured sizes 
for the Coma cluster objects are: CcGV1 = 264.6 $\pm$ 36.8, CcGV9a = 344.1 $\pm$ 47.9, CcGV9b = 
311.5 $\pm$ 43.3, CcGV12 = 152.3 $\pm$ 1.2, CcGV18 = 205.8 $\pm$ 28.6, CcGV19a = 208.8 $\pm$ 29.1, 
CcGV19b = 99.4 $\pm$ 13.8 pc.

\subsubsection{Photometry}

In addition to providing size estimates for our CSSs we have also obtained new, or reanalysed existing, 
imaging data for each CSS. Briefly, this photometry includes the optical HST images used to select the 
CSSs, new SOAR/Goodman U, B, V, $\&$ R images obtained for several southern hemisphere AIMSS 
CSSs, as well as reanalysed SDSS DR9 \citep{SDSSDR9} u, g, r, i, $\&$ z  photometry for equatorial and 
northern hemisphere objects within the SDSS footprint. Where possible we have also reanalysed archival 
2MASS \citep{Skrutskie06}, HAWK-I \citep{HAWK-I}, and NEWFIRM \citep{NEWFIRM} IR images for each 
CSS. Where the data required reduction (i.e. the SOAR/Goodman, HAWK-I, and NEWFIRM data) we 
made use of standard IRAF routines to carry out bias subtraction, flat fielding, and image co-addition. Zero 
points for the IR data were set using 2MASS stars located within the HAWK-I and NEWFIRM fields-of-view. 
For the Goodman data we made use of standard star fields (from \citealt{Landolt09}) observed at similar 
airmass, immediately after the science target to provide zero points accurate to $<$0.03 mag in all bands.

For all analyses we proceeded in a similar manner to that described in Section \ref{sec:halflight}. We first
downloaded the calibrated frames, used IRAF/ELLIPSE or a median subtraction to remove the large scale
host galaxy light, determined a curve-of-growth magnitude for the CSS, then applied a correction for 
foreground extinction based on \citet{Schlafly11}. We find that the background subtraction and
reanalysis is particularly vital for SDSS photometry, where the catalogued photometry frequently suffers
from catastrophically under or over-estimated magnitudes and always underestimates errors for CSSs 
near to larger galaxies (frequently providing errors of $<$ 0.01 mag for u band photometry of faint CSSs).

\subsection{Stellar Mass Determination}
\label{Sec:StellMass}
To determine stellar mass estimates for the CSSs, both for our AIMSS discovered objects and the objects compiled into 
our master catalog, we use a modified version of the stellar mass estimation code first presented in \cite{Kannappan07} 
and later updated in \citet{Kannappan13}. Briefly, the code fits photometry from the Johnson-Cousins, Sloan, and 2MASS 
systems with an extensive grid of models from \citet{BruzualCharlot} assuming a Salpeter initial mass function (IMF). 
Each collection of input CSS photometry is fitted by a grid of two-SSP, composite old+young models with ages from 5~Myr 
to 13.5~Gyr and metallicities from Z = 0.008 to 0.05. This range of age and metallicity is sufficient to adequately cover 
those displayed by all of our CSS types, from YMCs to ancient GCs. The derived stellar mass is determined by the median 
and 68$\%$ confidence interval of the mass likelihood distribution binned over the grid of models. Following 
the procedure used in \citet{Norris&Kannappan11} we rescale the derived stellar masses by a factor of 0.7 in order to match 
the ``diet" Salpeter IMF of \citet{Bell01}. We do this in order to make our stellar mass estimates more consistent with a 
Kroupa IMF which appears to be a better fit to observational data than Salpeter for both GCs \citep{Strader11a} and relatively
low mass early type galaxies \citep[those with $\sigma_{\rm e}$ $\sim$ 90 kms$^{-1}$;][]{ATLAS3DXX}. 
Therefore as the stellar masses of our GC, UCD, and cE sample overlap with, and transition between the stellar masses 
of GCs and low mass early type galaxies, a Kroupa IMF would appear to be the most logical choice of IMF to apply.

\begin{figure} 
   \centering
   \begin{turn}{0}
   \includegraphics[scale=1.0]{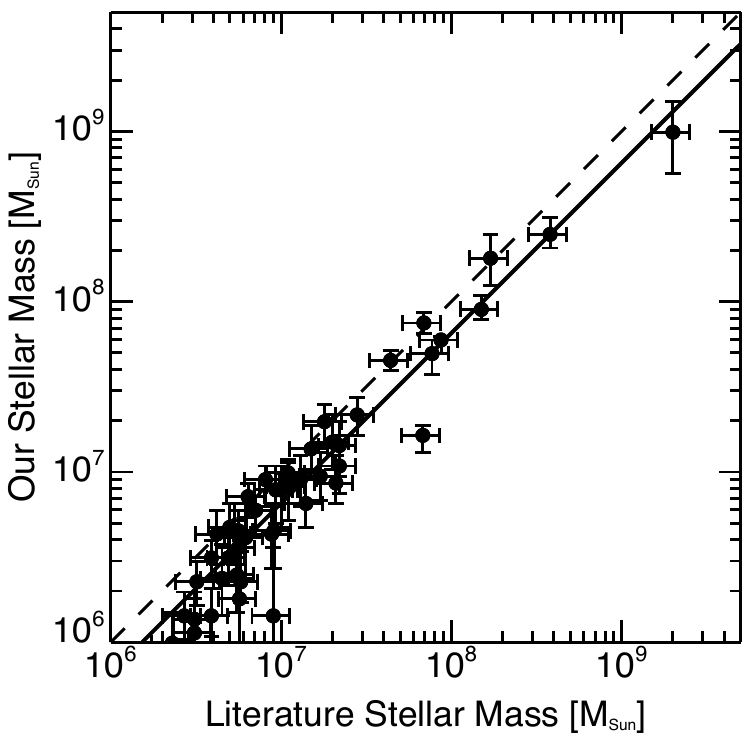}
   \end{turn} 
   \caption{Comparison of our derived stellar masses with literature values for 46 objects in common. The error 
   bars for our stellar masses are our errors derived using the procedure outlined in Sec \ref{Sec:StellMass}, while the error 
   bars for the literature data are purely illustrative (20$\%$ of measured values), as most literature analyses do not provide 
   errors. Systematic errors are not included but are $>$ 50$\%$ \citep[e.g.][]{Kannappan07}. The dashed line is the one-to-one relation, 
   while the solid line is the best fit relation for the data. Our stellar mass estimates are on average 65$\%$ of the literature ones,
   as expected given our assumption of a Kroupa-like IMF compared to a Salpeter IMF for most literature measurements.}   
   \label{fig:StellarMass_Comparison}
\end{figure}

Figure \ref{fig:StellarMass_Comparison} shows our derived stellar masses versus those from the literature (calculated
using a range of techniques including SSP fitting and single band M/L ratios) for a sample of 46 objects in common. The 
agreement between the different mass estimates is remarkably good, considering the inhomogeneous nature of the input 
photometry, and the different approaches used to estimate stellar mass in the literature. Our
stellar mass measurements are systematically lower, with ours on average being 65$\%$ of the literature values, almost
exactly as expected given that our stellar masses aim to be Kroupa-like, and most literature measurements are
made assuming a Salpeter IMF. There is also some evidence for a tendency for our stellar masses to be even lower than
expected when compared to the literature ones for M$_{\star}$ $<$ 5$\times$10$^6$ M$_{\odot}$. 
However, this only affects a handful of objects in the comparison, and at most 3 of the sample of 28 new objects presented 
here. The level of the divergence is also within the typical factor of 2 systematic error between different mass estimations.

\subsection{Classifying Host Galaxy Environments}

Until recently almost all confirmed compact stellar systems were discovered in massive galaxy clusters, leading to the belief 
that the cluster environment could be responsible for forming such systems \citep[e.g. forming UCDs by the ``threshing" of 
nucleated galaxies by cluster potentials:][]{Bekki01}. However, in recent years several CSSs located in field and group 
environments have been found \citep[e.g.][]{Hau09,Norris&Kannappan11,Huxor13}, indicating that cluster environments are 
not essential for CSS formation.

We deliberately did not use environment as a selection factor in choosing CSS candidates for spectroscopic observations, in 
order to ensure that we did not bias our selection in favour of high-density environments. However, after observing and 
confirming the nature of our CSSs we then made an (admittedly crude) estimate of the environments of their 
host galaxies. In general to classify the environment of the CSSs we again make use of the 2MASS All Sky Redshift Survey 
group catalog of \citet{Crook07}. To make a rough classification into field, group, and cluster environments we use the number 
of galaxies found in the same structure as the host galaxy, as found in the \citet{Crook07} catalog. Then using agreed classifications 
in the literature (i.e. that the Fornax cluster is a cluster, that NGC~3923 is in a group) we define the limits between environments 
as follows: field environments have 10 or fewer members, groups have more than 10 members and fewer than 40, clusters have 
more than 40 members.

Several AIMSS galaxies are not in the \citet{Crook07} catalog, so in these cases (NGC~0034, NGC~0821, NGC~1172, NGC~3115, 
and ESO383-G076) we base their environmental classification on literature determinations. We also reclassify NGC~4546 and 
the Sombrero galaxy as field galaxies, in contrast to the \citet{Crook07} determination that these are members of the Virgo 
cluster, despite their lying at least 3~Mpc from the Virgo cluster centre.

While admittedly very crude, this classification does at least allow us to demonstrate that compact
stellar systems of all masses are found associated with galaxies in a wide variety of environments, from very isolated galaxies 
such as NGC~4546, NGC~3115, or the Sombrero, to massive galaxy clusters. This should not be a surprising finding given 
the existence of the prototypical cE, M32, in a small group environment.

Our classifications are also in general in reasonable agreement with a more physical classification based on 
group central galaxy stellar mass to halo mass ratio, which is not applicable to all our sample galaxies because of missing stellar
masses for the relevant group dominating galaxies. Using this alternative approach, the stellar masses of
NGC~4546, NGC~3115, and the Sombrero (all considered dominant in their local environments) --- lie in the range expected 
for group centrals in halos near or just above the field-to-group transition at halo mass $\sim$1.3$\times10^{12}$ M$_\odot$ 
where galaxy formation efficiency peaks \citep{Leauthaud12}, but well below the group-to-cluster transition at halo mass 
$\sim$3$\times10^{13}$ M$_\odot$ where cluster quenching processes take over \citep{Robotham06}.  In particular, these 
transitional halo mass scales correspond to central galaxy stellar masses of $\sim$3$\times10^{10}$ M$_\odot$ and 
$\sim$1.3$\times10^{11}$ M$_\odot$ \citep[e.g.][]{Behroozi13}, while we measure stellar masses of 
$\sim$2.7$\times10^{10}$ M$_\odot$, $\sim$6.4$\times10^{10}$ M$_\odot$, and $\sim$8.2$\times10^{10}$ M$_\odot$
for NGC4546, NGC3115, and the Sombrero, respectively. As all three galaxies have early type morphology, considerable 
hierarchical merging has very likely occurred and may be involved in the creation of the CSSs, but physical processes specific 
to dense environments are less likely to be important.

\begin{table*}
\centering
\small
\begin{tabular}{|lllllll|}
\hline
Name 				& Distance		&	R$_{e}$			&	M$_{\rm V}$			& $\sigma_{\rm raw}$	& M$_{\star}$							&	Environment		\\
					& (Mpc)			&	(pc)				&	(mag)				& (km\,s$^{-1}$)		& (M$_\odot$)							&					\\
\hline
\multicolumn{6}{|l}{\bf AIMSS Targets}&\\
NGC 0524-AIMSS1		& 24.0$\pm$2.3	&  39.9$\pm$3.8		& --12.59$\pm$0.21			& 31.7$\pm$3.5		&	4.96$^{+0.00}_{-0.22}$$\times10^7$		&	G				\\
NGC 0703-AIMSS1		& 82.8$\pm$17.2	& 164.7$\pm$28.4		& --14.97$\pm$0.49			& 20.7$\pm$7.2		&	3.13$^{+1.19}_{-0.76}$$\times10^8$		&	C				\\
NGC 0741-AIMSS1		& 81.3$\pm$17.3	& 311.7$\pm$55.0		& --17.60$\pm$0.43			& 86.2$\pm$4.7		&	5.96$^{+0.28}_{-0.53}$$\times10^9$		&	G				\\
NGC 0821-AIMSS1		& 22.4$\pm$1.8	&    6.5$\pm$0.5 		& --12.08$\pm$0.23 			& 	-				&      4.73$^{+3.12}_{-1.46}$$\times10^6$ 	&  	F				\\
NGC 0821-AIMSS2 		& 22.4$\pm$1.8	&    8.4$\pm$0.5 		& --11.06$\pm$0.23 			& 	-			 	&      7.50$^{+0.20}_{-0.23}$$\times10^6$ 	&  	F				\\
NGC 0839-AIMSS1		& 56.0$\pm$19.2	& 16.0$\pm$4.1		& --12.33$\pm$0.64			& 	-				&	5.69$^{+3.75}_{-1.75}$$\times10^6$		&	F				\\
NGC 1128-AIMSS1		& 100.0$\pm$16.4	& 76.0$\pm$10.9		& --15.65$\pm$0.47			& 63.9$\pm$5.8		&	7.50$^{+0.15}_{-0.18}$$\times10^8$		&	G				\\
NGC 1128-AIMSS2		& 100.0$\pm$16.4	& 484.8$\pm$69.2		& --17.86$\pm$0.38			& 58.1$\pm$3.2		&	4.73$^{+0.70}_{-0.80}$$\times10^9$		&	G				\\
NGC 1132-UCD1		& 99.5$\pm$16.9	& 84.3$\pm$12.1		& --14.68$\pm$0.49			& 80.1$\pm$8.1		&	3.28$^{+0.85}_{-0.90}$$\times10^8$		&	F				\\
NGC 1172-AIMSS1		& 21.5$\pm$2.1	& 6.4$\pm$0.6			& --11.65$\pm$0.22			& 40.7$\pm$10.9		&	6.84$^{+3.51}_{-2.72}$$\times10^6$		&	F				\\
NGC 1172-AIMSS2		& 21.5$\pm$2.1	& 20.7$\pm$2.0		& --10.95$\pm$0.23			& 	-				&	1.72$^{+1.27}_{-0.73}$$\times10^6$		&	F				\\
Perseus-UCD13-AIMSS1	& 72.4$\pm$7.0	& 88.6$\pm$8.6		& --12.81$\pm$0.22			& 35.0$\pm$8.0		&	2.72$^{+1.21}_{-1.01}$$\times10^7$		&	C				\\
NGC 1316-AIMSS1		& 21.0$\pm$0.7	& 11.3$\pm$0.4			& --11.97$\pm$0.09			&	-				&	4.52$^{+4.10}_{-2.55}$$\times10^6$		&	C				\\
NGC 1316-AIMSS2		& 21.0$\pm$0.7	& 8.3$\pm$0.3			& --10.42$\pm$0.12			&	-				&	1.57$^{+0.92}_{-0.67}$$\times10^6$		&	C				\\
NGC 2768-AIMSS1		& 22.4$\pm$2.6	& 6.4$\pm$0.7			& --12.07$\pm$0.24			& 38.1$\pm$4.6		&	5.43$^{+4.01}_{-2.00}$$\times10^6$		&	F				\\
NGC 2832-AIMSS1		& 98.6$\pm$16.7	& 46.4$\pm$6.7		& --14.78$\pm$0.39			& 111.3$\pm$11.0		&	2.37$^{+0.90}_{-0.81}$$\times10^8$		&	F				\\
NGC 3115-AIMSS1		& 9.00$\pm$0.3	& 8.6$\pm$0.4			& --11.27$\pm$0.12 			& 36.9$\pm$1.9		&	1.09$^{+0.28}_{-0.37}$$\times10^7$		&	F				\\
NGC 3268-AIMSS1		& 39.8$\pm$2.8	& 25.6$\pm$1.9		& --12.68$\pm$0.18			&	-				&	3.43$^{+1.09}_{-1.16}$$\times10^7$		&	C				\\
NGC 3923-UCD1		& 21.3$\pm$2.9	& 12.3$\pm$0.3		& --12.43$\pm$0.28			& 33.0$\pm$2.1		&	1.97$^{+0.51}_{-0.61}$$\times10^7$		&	G				\\
NGC 3923-UCD2		& 21.3$\pm$2.9	& 13.0$\pm$0.2		& --11.93$\pm$0.28			& 23.1$\pm$3.6		&	6.53$^{+2.49}_{-1.80}$$\times10^6$		&	G				\\
NGC 3923-UCD3		& 21.3$\pm$2.9	& 14.1$\pm$0.2		& --11.29$\pm$0.29			& 15.5$\pm$3.6		&	2.37$^{+1.22}_{-0.40}$$\times10^6$		&	G				\\
NGC 4350-AIMSS1		& 16.5$\pm$0.8	& 15.4$\pm$0.1		& --12.16$\pm$0.15			& 25.5$\pm$9.0		&	1.57$^{+0.60}_{-0.53}$$\times10^7$		&	C				\\
NGC 4546-AIMSS1		& 13.1$\pm$1.3	& 25.5$\pm$1.3		& --12.94$\pm$0.20			& 21.8$\pm$2.5		&	3.59$^{+0.73}_{-0.99}$$\times10^7$		&	F				\\
NGC 4565-AIMSS1		& 16.2$\pm$1.3	& 17.4$\pm$1.4		& --12.37$\pm$0.20 			& 13.8$\pm$8.1		&	8.19$^{+0.42}_{-0.20}$$\times10^6$		&	C				\\
NGC 4621-AIMSS1		& 14.9$\pm$0.5	& 10.2$\pm$0.4		& --11.85$\pm$0.07			& 33.9$\pm$4.4		&	1.64$^{+0.43}_{-0.34}$$\times10^7$		&	C				\\
M60-UCD1			& 16.4$\pm$0.6	& 27.2$\pm$1			& --14.18$\pm$0.09			& 61.6$\pm$1.8		&	1.80$^{+0.18}_{-0.23}$$\times10^8$		&	C				\\
ESO 383-G076-AIMSS1 	& 162.8$\pm$15.6	& 86.1$\pm$7.6		& --15.52$\pm$0.25			&	-				&	4.96$^{+1.28}_{-1.20}$$\times10^8$		&	C				\\
NGC 7014-AIMSS1		& 58.6$\pm$4.2	& 329.8$\pm$23.6		& --15.17$\pm$0.16			& 20.6$\pm$6.3		&	2.99$^{+0.95}_{-1.02}$$\times10^8$		&	G				\\
					&				& 					& 						&					&									&					\\

\multicolumn{6}{|l}{\bf Reobserved objects, serendipitous observations, or objects with reanalysed photometry}&\\
Fornax-UCD3			& 20.0$\pm$1.4	& 86.5$\pm$6.2		& --13.45$\pm$0.10$^\diamond$	&  	27.4$\pm$5.1		&	4.96$^{+1.28}_{-1.02}$$\times10^7$		&	C			\\
NGC 2832-cE			& 98.6$\pm$16.7	& 375.3$\pm$54.4		& --17.77$\pm$0.34				& 	100.5$\pm$3.0		&	2.27$^{+0.59}_{-0.47}$$\times10^9$		&	F			\\
NGC 2892-AIMSS1 		& 97.7$\pm$16.6	&  580.9$\pm$85.0 		& --18.88$\pm$0.37 				& 137.5$\pm$3.7 		&    	1.09$^{+0.11}_{-0.14}$$\times10^{10}$	&  	F			\\
NGC 3268-cE1/FS90 192	& 39.8$\pm$2.8	& 299.9$\pm$21.9		& --15.92$\pm$0.16				&	36.8$\pm$15.0		&	1.30$^{+0.41}_{-0.11}$$\times10^8$		&	C			\\
Sombrero-UCD1		& 9.00$\pm$0.1	& 14.7$\pm$1.4		& --12.31$\pm$0.06$^\diamond$	&	 31.9$\pm$2.9		&	1.64$^{+0.43}_{-0.40}$$\times10^7$		&	F			\\
M59cO	 			& 14.9$\pm$1.1	& 35.2$\pm$1.2		& --13.43$\pm$0.09				&   	29.0$\pm$2.5		&	7.49$^{+0.11}_{-0.10}$$\times10^7$		&	C			\\
ESO 383-G076-AIMSS2	& 162.8$\pm$15.6	& 652.2$\pm$57.5		& --17.35$\pm$0.25				& 	87.8$\pm$8.0		&	2.60$^{+0.53}_{-0.53}$$\times10^9$		&	C			\\
					&				& 					& 							&					&									&				\\

\multicolumn{6}{|l}{\bf YMC Sample (all literature values except stellar masses)}&\\

NGC 0034-S$\&$S1		& 85.1$\pm$17.2	& 39.0$\pm$7.8$^\star$ 		& --15.36$\pm$0.40		&	-				&	3.13$^{+4.73}_{-1.94}$$\times10^7$		&	F				\\
NGC 0034-S$\&$S2		& 85.1$\pm$17.2	& 31.9$\pm$6.4$^\star$		& --14.70$\pm$0.40		&	-				&	2.37$^{+2.82}_{-1.47}$$\times10^7$		&	F				\\
NGC 1316-G114		& 21.0$\pm$0.7	& 42.1$\pm$2.6			& --12.81$\pm$0.12		&	42.1$\pm$2.8		&	2.72$^{+0.71}_{-1.09}$$\times10^7$		&	C				\\
NGC 7252-W3			& 67.3$\pm$17.4	& 17.7$\pm$4.4$^\dagger$	& --16.30$\pm$0.51		&	45.5$\pm$5.2		&	1.25$^{+1.74}_{-0.89}$$\times10^8$		&	F				\\
NGC 7252-W6			& 67.3$\pm$17.4	& 5.1$\pm$1.3$^\dagger$		& --14.50$\pm$0.51		&	-				&	2.37$^{+2.36}_{-1.47}$$\times10^7$		&	F				\\
NGC 7252-W26		& 67.3$\pm$17.4	& 11.9$\pm$3.0$^\dagger$	& --13.75$\pm$0.51		&	-				&	1.37$^{+1.24}_{-0.82}$$\times10^7$		&	F				\\
NGC 7252-W30		& 67.3$\pm$17.4	& 8.3	$\pm$2.1$^\dagger$		& --14.68$\pm$0.51		&	27.5$\pm$2.5		&	2.72$^{+2.71}_{-1.74}$$\times10^7$		&	F				\\

\hline
\end{tabular}

\caption[Properties UCDs]{Physical Properties of the AIMSS objects. Distances for the host galaxies are surface brightness fluctuation 
distances listed in NED\footnote{http://ned.ipac.caltech.edu/} where available. Where no distances are available in the literature we 
assume Hubble flow distances assuming H$_{\rm 0}$ = 70 km\,s$^{-1}$Mpc$^{-1}$. Magnitude and size errors include the distance
uncertainty. Because of limited photometry no reliable stellar mass estimates were possible for NGC~1172 AIMSS 1 and 2. \\
$^\diamond$ Errors calculated by assuming photometric uncertainty of 0.05 mag combined with the measured distance uncertainty.\\
$^\star$ Sizes from \citet{Schweizer07} with errors computed assuming 0.4 mag distance modulus uncertainty to NGC~34. \\
$^\dagger$ Sizes from \citet{Bastian13} with errors computed assuming 0.5 mag distance modulus uncertainty to NGC~7252.
}
\label{properties} 
\end{table*} 

\section{Results}

Table \ref{properties} provides the measured properties of the AIMSS targets, as well as the six objects reobserved as a consistency 
check, and the comparison sample of seven YMCs.

\subsection{Luminosity -- Effective Radius}

\begin{figure*} 
   \centering
   \begin{turn}{0}
   \includegraphics[scale=1.0]{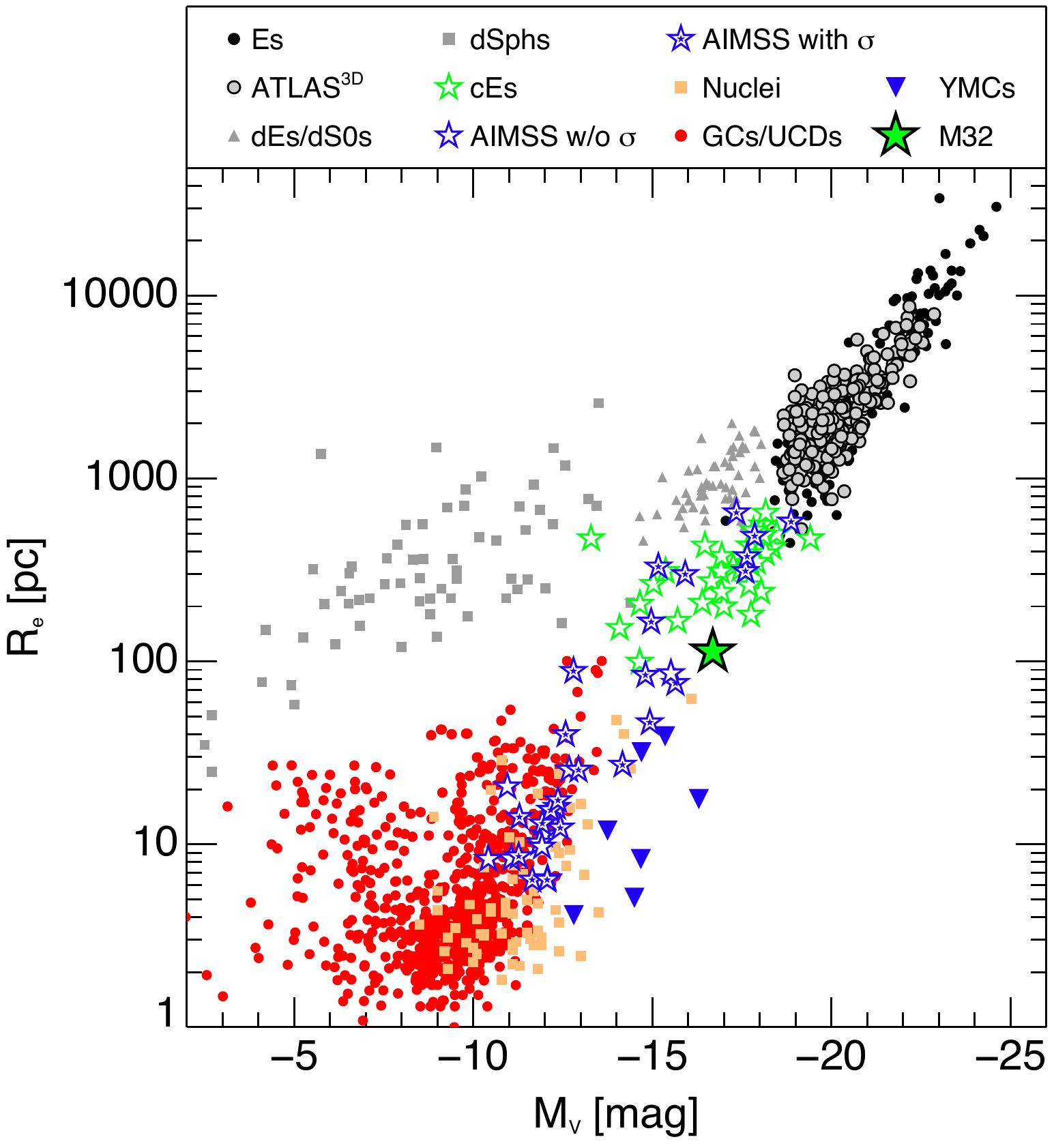}
   \end{turn} 
   \caption{Effective radius R$_{\rm e}$ vs absolute V-band magnitude M$_{\rm V}$ for dynamically hot stellar systems from our 
   master compilation. Blue stars are our observations and are filled where we have successfully measured the velocity dispersion 
   of the object. M32 is indicated by its own symbol and labelled. It is clear that the group of six (including one literature Coma cE) 
   non-YMC objects with M$_{\rm V}$ $<$ --14 and R$_{\rm e}$ $<$ 100pc lie offset significantly from the more massive previously 
   known UCDs such as Virgo-UCD7 and Fornax-UCD3, which are smaller than M32.}
   \label{fig:Mv_Reff}
\end{figure*}

Figure \ref{fig:Mv_Reff} shows the location of the AIMSS sample (blue stars) in size--luminosity space relative to other stellar systems, 
including GCs and previously known UCDs (red dots), galaxy nuclei (orange squares), Young Massive Clusters (blue triangles), cEs 
(green open stars), dSphs (grey filled squares), dEs and dS0s (grey filled triangles), elliptical galaxies (filled black dots), Es and S0s 
from the ATLAS$^{3D}$ (grey dots), and the prototypical cE M32 indicated with a symbol of its own. 

From Figure \ref{fig:Mv_Reff} it is clear that most AIMSS objects occupy the previously defined region of parameter space for UCDs 
and cEs. However, of immediate interest is the presence of a population of \textbf{six} objects (five AIMSS objects plus one previously 
known Coma cluster cE) that are considerably brighter at fixed size than previously known UCDs (with $\sim$ 30 $<$ R$_{\rm e}$ [pc] 
$\la$ 100, and M$_{\rm V}$ $<$ --14.0). The smallest of these six objects (M60-UCD1, M$_{\rm V}$ $\sim$ --14, R$_{\rm e}$ 
$\sim$30pc) has been previously described in \cite{Strader13} as the ``Densest Galaxy". Our newly discovered objects indicate 
M60-UCD1 is the first of a population of unusually compact and luminous (even for UCDs) stellar systems. The six objects lie within 
a region to the right (i.e. more luminous) side of the usual UCD trend that had previously only been inhabited by much younger and 
hence more luminous YMCs (see Section \ref{Sec:Mass_Re}) and by the nuclei of galaxies. The new objects also appear to extend 
the apparent hard limit on the bright side of the elliptical galaxy size--luminosity trend all the way down to the star cluster mass regime.

These new objects are found in all environments, with two found in the field (NGC~1132-UCD1 and NGC~2832-AIMSS1), one found 
in a group (NGC~1128), and three found in clusters (CcGV19b, NGC~4649-AIMSS1, ESO383-G076-AIMSS1). This diversity should 
not be too surprising as M32 itself is found in a low-N, field-like group, proving that cluster environments are not essential for forming 
the densest stellar systems.

\begin{figure} 
   \centering
   \begin{turn}{0}
   \includegraphics[scale=0.625]{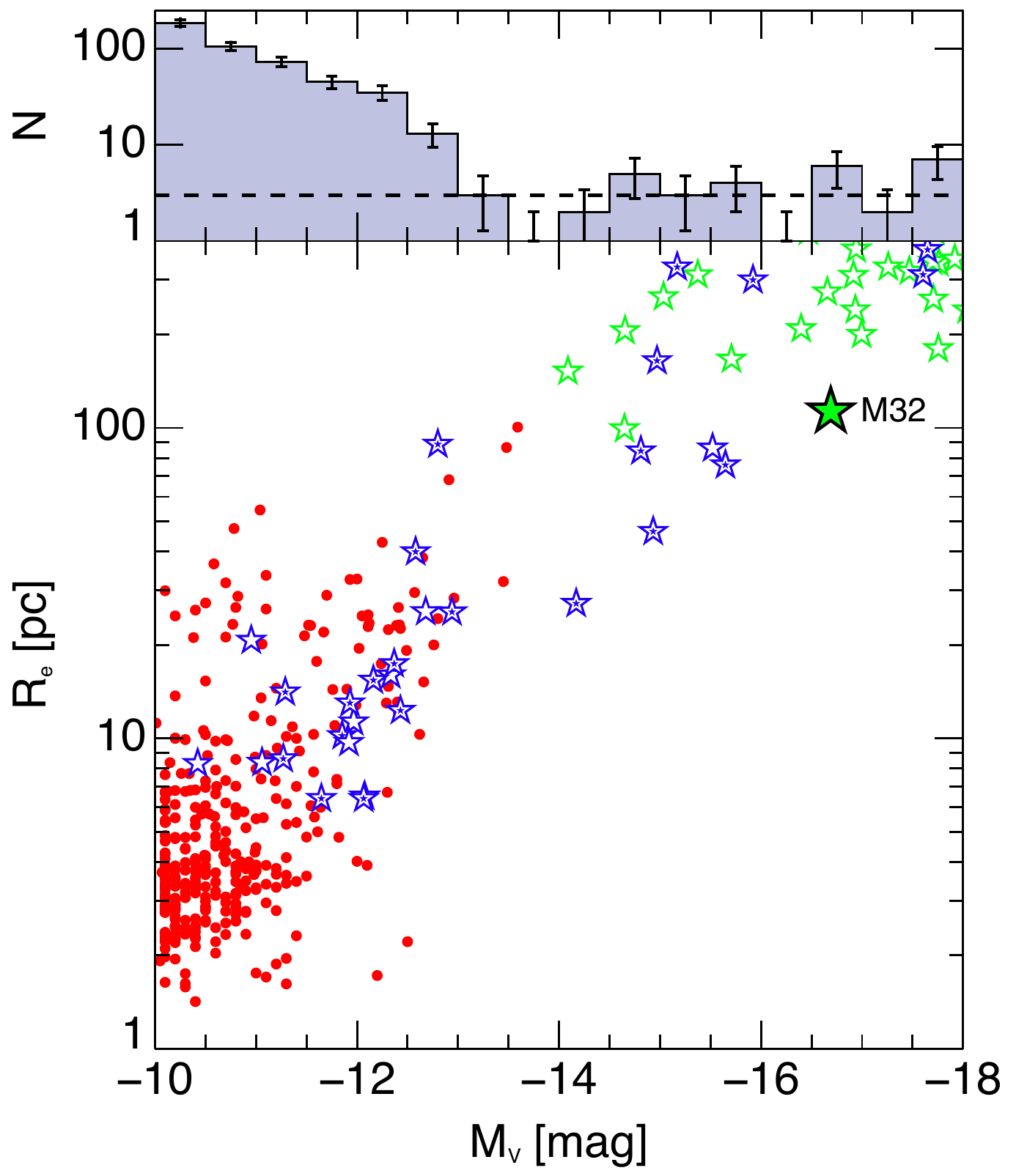}
   \end{turn} 
   \caption{$\bf{Lower~Panel:}$ Effective radius vs absolute V-band magnitude for all known massive GCs, UCDs, and cEs with 
   \mbox{--10} $<$ M$_{\rm V}$ $<$ --18 and R$_{\rm e}$ $<$ 400 pc. $\bf{Upper~Panel:}$ Histogram of the M$_{\rm V}$ values of 
   the selected objects, with the statistical 1$\sigma$ uncertainties shown by the error bars. The dashed line shows the median 
   number of objects (three) per bin for M$_{\rm V}$ $<$ --13. It is clear that there is a drop off in the number of objects for M$_{\rm V}$ 
   $<$ --13.}
   \label{fig:Mv_Reff_Hist}
\end{figure}

Another observation to be made from Figure \ref{fig:Mv_Reff} is the apparent rarity of compact stellar systems more luminous than 
M$_{\rm V}$ = --13. This observation is examined in more detail in Figure \ref{fig:Mv_Reff_Hist}, where all known massive GCs, 
UCDs, and cEs with --10 $<$ M$_{\rm V}$ $<$ --18 and R$_{\rm e}$ $<$ 400 pc are plotted. The top panel of this figure shows the 
histogram of the magnitudes of these objects, with a dashed horizontal line denoting the median number of objects per bin with 
M$_{\rm V}$ $<$ --13. There appears to be a roughly constant number of objects brighter than M$_{\rm V}$ = --13, but fainter than 
this value, the number of objects increases dramatically. Of course the sample being examined here is in no way homogeneous, 
having been built from many disparate surveys, and is therefore not a well defined statistically complete sample. However, it is also 
clear that brighter objects at fixed size are observationally $\emph{simpler}$ to find, so the observed drop-off in CSS frequency fainter 
than M$_{\rm V}$ $<$ --13 is not likely to be due to simple selection effects in the surveys used to build the sample. In fact many of 
the input literature samples were specifically designed to spectroscopically observe all objects irrespective of size brighter than 
magnitude limits of around M$_{\rm V}$ $\sim$ --15.5 \cite[e.g.][]{Jones06}, further indicating that the observed drop off is likely real. 
It is also interesting that M$_{\rm V}$ $>$ --13 is exactly the upper limit for objects formed in star cluster like processes suggested 
by \cite{Norris&Kannappan11} on the basis of statistical arguments about the globular cluster luminosity function, as first suggested 
by \cite{Hilker09}.

\subsection{Stellar Mass -- Effective Radius}
\label{Sec:Mass_Re}

\begin{figure*} 
   \centering
   \begin{turn}{0}
   \includegraphics[scale=1.0]{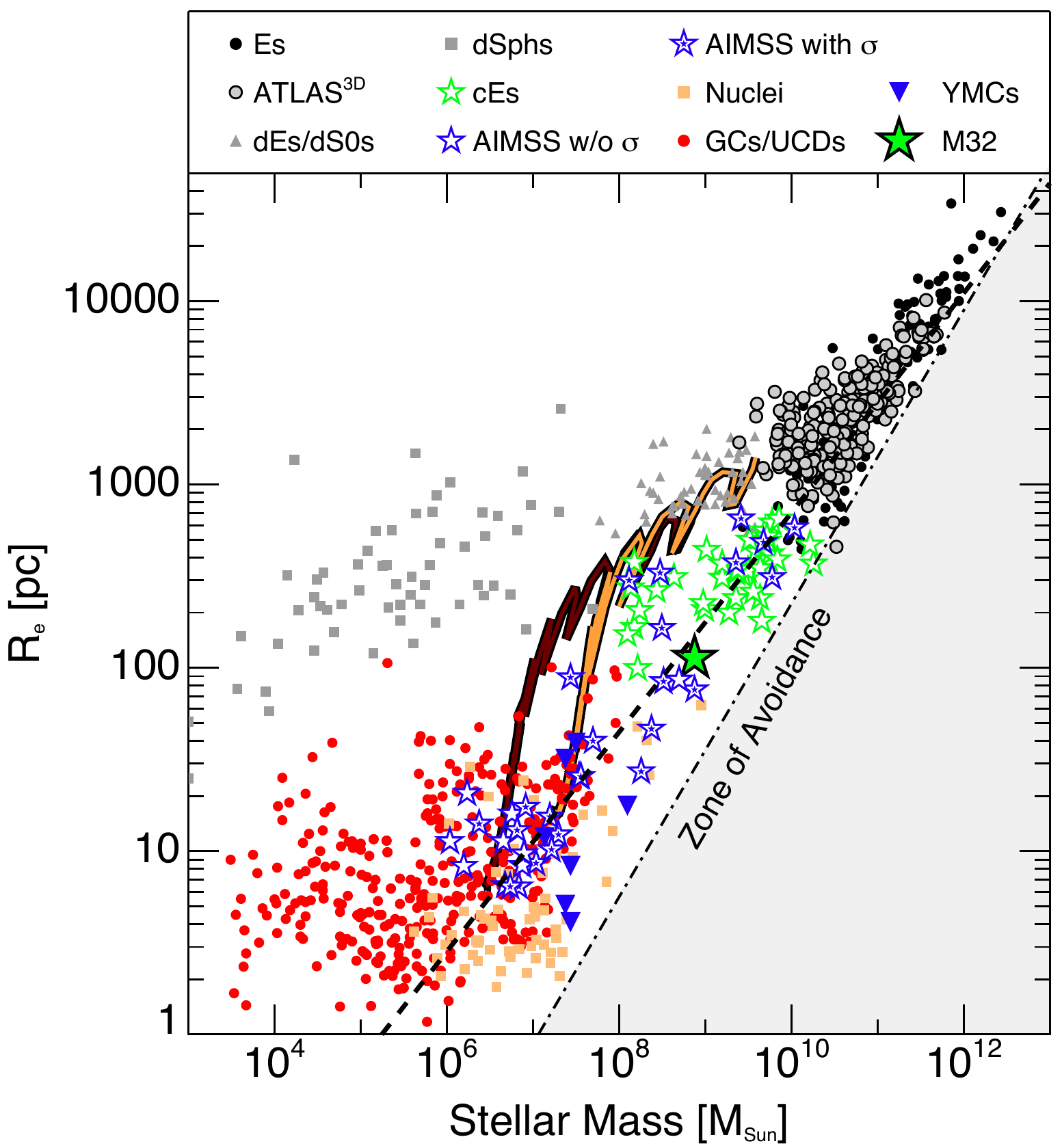}
   \end{turn} 
   \caption{Effective radius R$_{\rm e}$ vs stellar mass for compact stellar systems and comparison samples. Symbols are the same 
   as those defined in Figure \ref{fig:Mv_Reff}. The dot-dash line is the by-eye fit to the edge of the edge of the elliptical galaxies, cEs 
   and dE nuclei, as determined by \citet{Misgeld&Hilker11}, having the form R$_{\rm eff}$(M) $>$ 2.24 $\times$ 10$^{-6}$ M$_{\star}^{4/5}$ 
   pc M$_\odot^{4/5}$. The orange and brown solid lines show the simulated evolution of two nucleated dEs as they are stripped by the 
   potential of a larger galaxy from \citet[][their simulations 3 and 17]{Pfeffer13}. The dashed line is the fit to bright ellipticals, compact 
   ellipticals, bulges, and UCDs from \citet{Dabringhausen08}. In this plot the separation between the unusually dense objects (mostly to 
   the right of the dashed line) and the previously known UCDs (near and to the left of the dashed line) is even more clearly demonstrated.  
   }
   \label{fig:Mass_Reff}
\end{figure*}

Returning to the question of the six unusually bright objects in the M$_{\rm V}$ -- R$_{\rm e}$ parameter space, one explanation for 
their location is that, like YMCs, these objects are younger, and hence brighter than the generally old UCD population at the same 
stellar mass. This possibility is a valid concern as some cEs are observed to have intermediate stellar ages (e.g. \citealt{Schiavon04}, 
\citealt{Chilingarian09}, \citealt{Miner11}, \citealt{Huxor11b}), although this may be due to a ``frosting" of recent star formation and not 
a dominant mass component. As we currently lack suitable spectroscopic estimates of age for many of our objects we therefore turn 
to the stellar mass estimates to address this question.

Figure \ref{fig:Mass_Reff} convincingly demonstrates that age differences are not responsible for the observed offsets. In the stellar 
mass -- R$_{\rm e}$ plane the gap between the six unusual objects and the main UCD track (as defined by the dashed-line fit to bright 
ellipticals, compact ellipticals, bulges and UCDs from \citet{Dabringhausen08}) is decreased, but five of the six remain to the right of 
the main UCD track. The behaviour of the YMCs in this plane is also instructive, as these truly young clusters, which in the M$_{\rm V}$ 
-- R$_{\rm e}$ plane all lie far to the right of the main UCD trend, are now mostly consistent with the general trend, indicating that they 
were offset due to their youthful luminosity.

The location of the unusually bright objects also demonstrates that three of the objects (NGC~1128-AIMSS1, NGC~1132-UCD1, 
ESO383-G076-AIMSS1) are amongst the best analogues in terms of mass and size for M32 yet found. 

We over plot on Figure \ref{fig:Mass_Reff} the evolutionary tracks (solid orange and brown lines) of two nucleated dE galaxies as they 
are tidally stripped from the simulations of \citet{Pfeffer13}. The simulations are numbers 3 (brown) and 17 (orange) from \citet{Pfeffer13}. 
They are both of dE,N galaxies on elliptic orbits with apocenter of 50 kpc and pericenter of 10 kpc around a cluster centre which has 
properties chosen to match M87 in the Virgo cluster. Simulation 3 originally has a nucleus with R$_{\rm e}$ = 4 pc and M$_{\rm V}$ 
= --10, Simulation 17 initially has a nucleus with R$_{\rm e}$ = 10 pc and M$_{\rm V}$ = --12. Both are simulated for a total of 4.2 Gyr.  
We use these simulations to stand in for simulations of any nucleated dwarf galaxies undergoing stripping, as at present very few 
simulations of the stripping of later type dwarfs have been carried out, but we expect that the stripping of other dwarf galaxy types should 
produce reasonably similar results. The simulations of \citet{Pfeffer13} demonstrate that the remnants of the stripping of dE,Ns can 
resemble almost all massive GCs and UCDs, even the most extended (R$_{\rm e}$ $\sim$ 100pc) and massive (M $\sim$ 10$^8$ 
M$_\odot$) UCDs such as Fornax-UCD3, Virgo-UCD7, and Perseus-UCD13. However, it is also clear that these simulations cannot 
reproduce the properties of the unusually massive compact objects. To produce such objects by stripping requires the objects being 
stripped to be significantly more massive initially, making their likely progenitors true ellipticals, S0s, or bulged spiral galaxies.

A further interesting observation in Figure \ref{fig:Mass_Reff} is that despite the very high density of some of the new AIMSS objects, 
none of them (and no objects at all) have significantly violated the dot-dashed line into the region called the ``zone of avoidance" by 
\citet{Misgeld&Hilker11}. This region to the right of the dot-dash line is the equivalent of the ``zone of exclusion" that \citet{Burstein97} 
found to exist for early type galaxies. It therefore appears that there is a universal relation limiting the maximum stellar density an old 
dynamically hot stellar system may have, although this is not as simple as the limit being a constant mass surface density limit (see 
Sec \ref{Sec:MSurDens}). The existence of such a universal relation would be extremely interesting, especially considering the huge 
differences in composition and structure between apparently dark matter free GCs and dark matter dominated giant ellipticals. We 
discuss this topic in more detail in Section \ref{Sec:ZoA}.

In contrast to the further support for the existence of a ``zone of avoidance", the new AIMSS objects, along with other less massive but 
large GC-like objects (e.g. \citealt{Brodie11}, \citealt{Forbes13}), and fainter cEs continue to weaken evidence for a suggested 
mass--size relation (see e.g. \citealt{Dabringhausen08}, \citealt{Murray09}, \citealt{Norris&Kannappan11}) for compact stellar systems. 
It is now clear that compact stellar systems can display a range in mass which varies by more than a factor of 100, at fixed size. This is 
more than a factor of 10 times the range seen for massive early-types. Observationally difficult to find fainter stellar systems with large 
radii are gradually being discovered and these fill in the space between classical GCs and dSphs. It seems possible that given time this 
region will be completely filled, further blurring the distinction between star clusters and galaxies.  

One final observation is that there are hints of a dichotomy in the cE population, between objects that appear to be a continuation of the 
elliptical galaxy population (the block centred at around 400pc and 5$\times$10$^9$ M$_\odot$), and a population possibly associated 
with dwarf galaxies (the tail dropping down from around 300 pc and 2$\times$10$^8$ M$_\odot$), we investigate this point further in 
Section \ref{Sec:MSurDens}.

\subsection{Stellar Mass -- Mass Surface Density}
\label{Sec:MSurDens}

\begin{figure*} 
   \centering
   \begin{turn}{0}
   \includegraphics[scale=1.0]{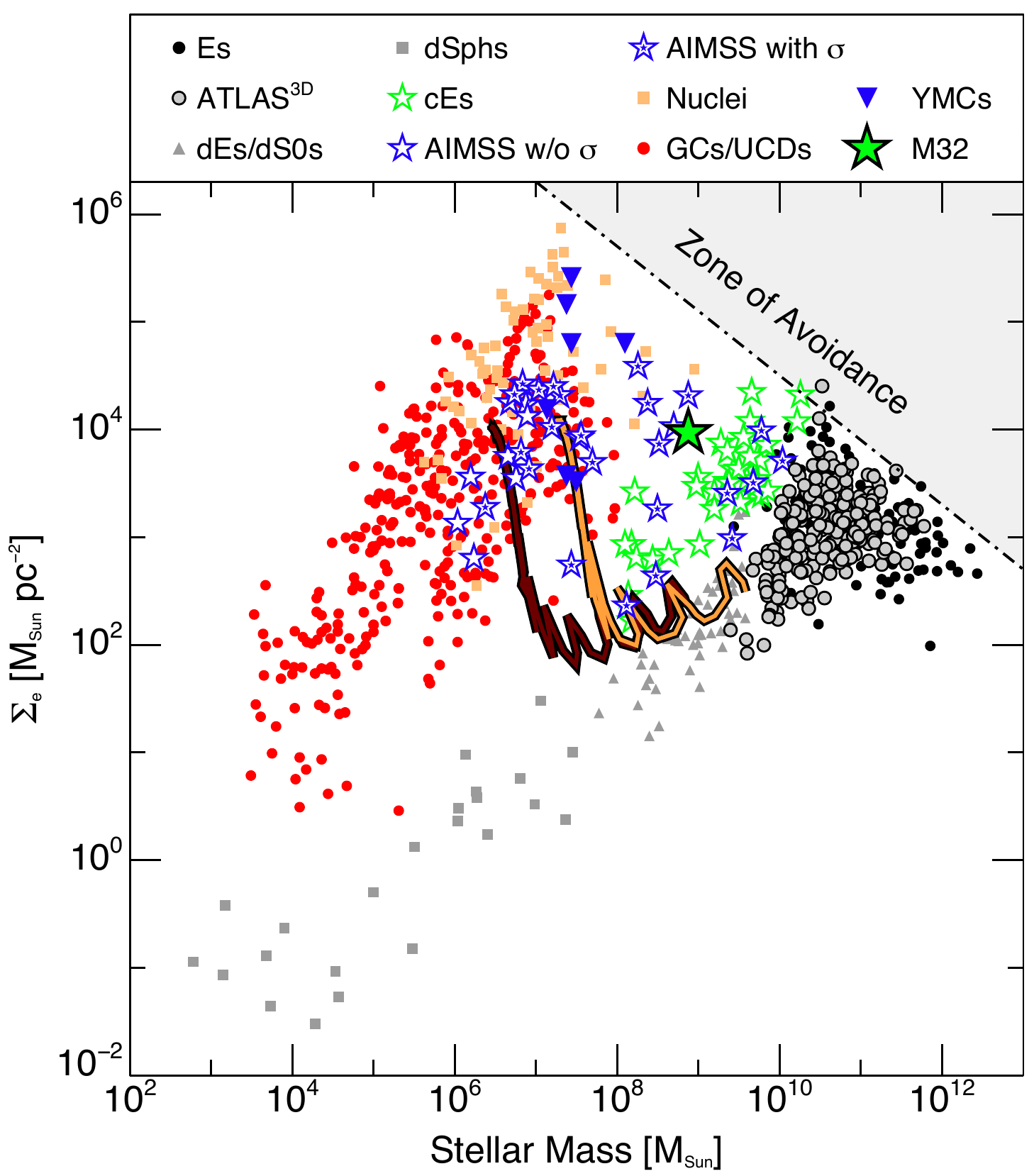}
   \end{turn} 
   \caption{Stellar mass vs. effective stellar mass surface density for dynamically hot stellar systems. Symbols are the same as those 
   defined in Figure \ref{fig:Mv_Reff}. The effective stellar mass surface density is calculated as $\Sigma_{\rm e}$ = M$_{\star}$ / 
   2$\pi$R$^2_{\rm e}$. The ``zone of avoidance" translated from that shown in Figure \ref{fig:Mass_Reff} is shown by the grey shaded 
   region. The solid orange and brown lines are the same simulations from \citet{Pfeffer13} as shown in Figure \ref{fig:Mass_Reff}. In 
   this plane it is even clearer that there is a separation between the densest objects discussed here and the majority of the compact
   stellar system population. }
   \label{fig:Mass_MassSurf}
\end{figure*}

Figure \ref{fig:Mass_MassSurf} shows the location of the various stellar systems in the stellar mass -- effective mass surface density 
(half the stellar mass divided by the area within the half-light radius) plane. This space is a mass vs average surface brightness plot, 
where the age dependence of the surface brightness is removed.

It is clear from this plot that several of our new AIMSS objects (including M60-UCD1; \citealt{Strader13}) lie in a region of parameter 
space where previously only M32, or the most massive galaxy nuclei (or bulges), were known to exist. It is also of note that the 
simulated stripped dE,Ns from \cite{Pfeffer13} also cannot reproduce the location of these dense stellar systems in the stellar mass -- 
effective mass surface density plane, again indicating a likely origin for most of these objects in more massive non-dwarf galaxies, if 
they are indeed the result of stripping. In fact, of this group only M60-UCD1 has properties that could be explained as being the result 
of the stripping of a dwarf galaxy (albeit on the massive end of the dwarf population), whereas all the rest must have resulted from the 
stripping of giant galaxies. 

It is also immediately obvious in Figure \ref{fig:Mass_MassSurf} that several AIMSS objects, and literature cEs, lie in a region of 
parameter space much closer to dwarf galaxies than to classical ellipticals, although the majority of cEs found to date are closer to
classical Es (with a division at $\sim$7$\times$10$^8$M$_\odot$). This observation may reflect a dichotomy in galaxy stripping. In a stripping 
scenario of UCD or cE formation, the stripping process simply removes the outer parts of galaxies to reveal the bound stellar structure 
within the galaxy, be it a nucleus, or for larger cEs potentially an entire galaxy bulge. Therefore the process should work for any galaxy 
with a central bound structure like dEs/dS0s, Es/S0s, or spirals, but not most dSph galaxies. This picture implies two types of resulting 
stripped object, one from the stripping of nucleated dwarfs and another form the stripping of bulged massive galaxies. It is also possible 
that some cEs are the result of dissipative merging like classical ellipticals (see e.g. \citealt{Kormendy12}).

In the stripping scenario a galaxy can move towards the upper left of Figure \ref{fig:Mass_MassSurf}, as the tightly bound stellar 
structure at the centre of the galaxy comes to dominate more and more of the total remaining galaxy. Therefore, a dwarf with a central 
nuclear star cluster similar to those on the plot will move away from the dwarf sequence towards the UCD region, as it is being stripped. 
This is precisely what the simulations by \cite{Pfeffer13} clearly show; that a dE,N when stripped will gradually evolve downward in the 
luminosity--size plot from the dwarf space, through the region inhabited by large UCDs (such as Fornax-UCD3 and Virgo-VUCD7), 
possibly even reaching sizes and luminosities indistinguishable from GCs. 

Figure \ref{fig:Mass_MassSurf} also demonstrates that the densest old stellar systems found in the Universe are galaxy nuclei. Our 
comparison sample of YMCs do reach similar densities, but after 10~Gyr of dynamical evolution and stellar mass loss their structures 
are likely to change dramatically.

\subsection{Stellar Mass -- Velocity Dispersion}
\label{Sec:MassSig}

\begin{figure*} 
   \centering
   \begin{turn}{0}
   \includegraphics[scale=1.0]{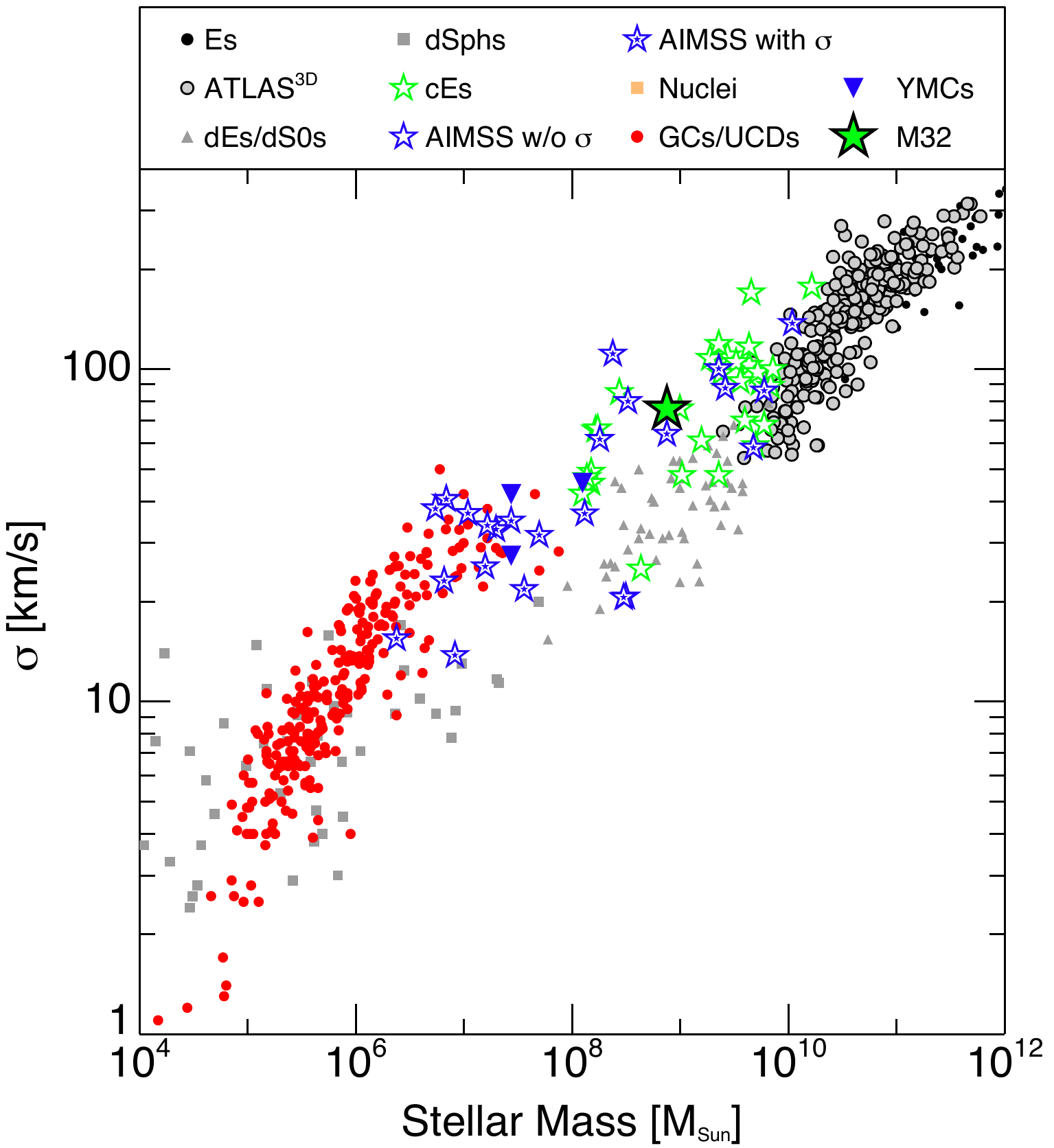}
   \end{turn} 
   \caption{Stellar mass vs. velocity dispersion for our various stellar systems. There is an obvious bifurcation between the star clusters 
   on one side, galaxies on the other, and CSSs on both loci. In this space tidal stripping tends to move objects towards the left (lower 
   mass) side of the plot at roughly constant velocity dispersion \citep{Bender92,Chilingarian09}. Therefore, most cEs and massive UCDs 
   are consistent with being stripped from objects originally 10--100 times larger than their current mass (mostly lower mass Es/S0s and 
   more massive dEs). However a possible second group of objects (with the most extreme examples being two AIMSS objects and one 
   literature cE with $\sigma\sim$ 20 km\,s$^{-1}$) with properties more like those of normal dEs also seem to exist, with these objects 
   perhaps being less severely stripped nucleated dwarf galaxies. }
   \label{fig:Mass_Sig}
\end{figure*}

Figure \ref{fig:Mass_Sig} shows the location of the various stellar systems in the stellar mass vs. velocity dispersion ($\sigma$) plane. It 
is clear from this figure that in agreement with \citet[][their Figure 5]{Forbes08} we find that there is a distinct separation between early-type 
galaxies and star clusters, with transitional objects such as massive UCDs and cEs bridging the gap between the two sequences at the high 
$\sigma$ end, and dSphs bridging the gap at the low $\sigma$ end. In the region between 10 and 70 km\,s$^{-1}$ there is a clear splitting 
of the plot into two tracks. In particular it is clear that the massive ellipticals display one slope, with a break at a central $\sigma$ of around 
105 km\,s$^{-1}$, followed by lower mass E's, S0s, dwarfs (dEs and dS0s on the plot) displaying a steeper slope which is closer to that of 
the GCs. However, the dEs/dS0s and more massive dSphs are offset from the GCs/UCDs towards higher masses, by a maximum factor of 
around 100 at 30 kms$^{-1}$. This offset is in good agreement with the M$_{\rm K}$ $\sim$ 5 offset between the two sequences determined 
by \citet{Forbes08}.

Also of interest in Figure \ref{fig:Mass_Sig} is the location of the cE population. The majority of the cEs lie offset to the left (lower mass) from 
the main galaxy trend, exactly as expected for objects that have been stripped down from initially larger normal galaxies. The magnitude of 
the offset (assuming central velocity dispersion is unaffected by the stripping process as claimed by \citealt{Bender92} and \citealt{Chilingarian09}) 
suggests that the cEs have lost up to 99$\%$ of their original mass. There is also evidence in this plot for the dichotomy we suggest exists in 
the cE population, with several of the AIMSS and literature cEs having velocity dispersions similar to those of massive dwarfs ($\sim$40-50 
km\,s$^{-1}$), while the bulk of the cE population has velocity dispersions more consistent with the elliptical sequence ($\sim$ 100 km\,s$^{-1}$).

\section{Discussion}

\subsection{Compact Stellar System Formation}

In the last few years it has become generally accepted that the UCD population is composite \cite[e.g.][]{daRocha11,Chilingarian11,
Norris&Kannappan11,Chiboucas11,Strader11b,Brodie11,Mieske13}, with both UCDs arising in star formation events (which also produce ``normal" 
star clusters such as GCs) as well as a population of objects resulting from the tidal stripping of galaxies. One piece of evidence for this duality 
comes from the normally close correspondence between the frequency and luminosity of UCDs and the total number of GCs in the host galaxy 
GC population \citep{Hilker09,Norris&Kannappan11,Mieske12}, combined with the observation that outliers exist which cannot be adequately 
explained by extrapolation of the GCLF \citep{Norris&Kannappan11}.

It is also the case that several of these objects, which cannot be explained by an extrapolation of the GCLF, have properties indicative of a 
stripped origin. For example, the UCD of NGC~4546 is found to be young ($\sim$3~Gyr) while its host galaxy is uniformly old ($\sim$10~Gyr). In 
addition, this UCD was found to counter rotate around the host galaxy, which is a ``smoking gun" of an accretion event. Penny et al. (submitted) 
also find that UCD13 in the Perseus cluster is likely to be the result of the stripping of a nucleated dwarf galaxy by NGC~1275, based on the UCD's 
colour, size, metallicity, velocity dispersion, dynamical mass, and proximity to NGC~1275. 

A further piece of evidence for the duality of UCD types comes from the observed kinematics of UCDs around M87. \citet{Strader11b} found that 
the position--velocity patterns of the UCDs near to M87 showed signatures of both radial and tangential orbits, as would be expected from stripped 
galaxies and surviving extended star clusters respectively. 

As already discussed, the simulations of \citet{Pfeffer13} clearly show that objects resembling Perseus-UCD13, Virgo-UCD7, Fornax-UCD3 and 
NGC~4546-UCD1 can be produced by the stripping of nucleated dwarf galaxies, with the largest UCDs ($\sim$R$_{\rm e}$ = 50--100 pc) being 
created when some of the stellar envelope of the dwarf is retained. However, the stripping of dwarf galaxies cannot explain the production of the 
new M32-like objects we present here, as the total stellar masses of dwarfs themselves are roughly the same as the final structures we seek to 
explain ($\sim$10$^{8}$--10$^{9}$ M$_\odot$), while dE/dS0 radii are a factor of 10 too large. Therefore, to create these new M32-like objects 
we must assume that either they are created by the stripping of bulges from spiral or early-type galaxies as suggested by e.g. \citet{Faber73}, 
\citet{Bekki01b}, \citet{Chilingarian09} and observed to occur in certain cases \citep[e.g.][]{Forbes03,SmithCastelli08,Huxor11b}, or that these 
objects are merely the low-luminosity extension of the true elliptical sequence \citep[e.g. ][]{Kormendy12}, or some combination of the two scenarios. 
We note that \cite{Chilingarian09} included some simulations of a tidally-stripped barred spiral galaxy in a Virgo cluster-like potential. They found 
that stripping leads to a strong increase in surface brightness (see their figure 1) similar to that predicted by \citet{Pfeffer13} for stripping of dE,N 
galaxies. They also confirmed the conclusion of \citet{Bender92} that tidal stripping moves objects to a lower stellar mass whilst leaving the velocity 
dispersion largely unchanged.  

An additional observation about the M32-like objects is that they are found in a range of environments which are very different to the local group 
where M32 is located. One is located near an isolated elliptical, another is in a group, and the last is in a dense cluster, further indicating the universal 
nature of the formation mechanism for this type of object. These new objects therefore carry the apparently paradoxical message that although they 
are rare, they are also ubiquitous, perhaps indicating that they are commonly created but short lived.

Having now determined that at least two types of process can create compact stellar systems, star formation related and galaxy interaction related, 
it remains to determine which objects were formed by which process. The close connection between UCD luminosity and GC system size, combined 
with the known properties of the GCLFs of galaxies, led \citet{Norris&Kannappan11} to suggest M$_{\rm V}$ $\sim$ --13 (M$_\star$ $\sim$ 7$\times$
10$^7$ M$_\odot$) as the upper limit of ``star cluster" type UCD formation. Above this mass limit all objects would be stripped nuclei or bulges, and 
below the limit a combination of both types would exist, with star cluster type UCDs increasingly dominating to lower masses. The suggested upper 
mass limit is close to the dividing mass found by \cite{Mieske13} (M$_\star$ $>$ 10$^7$ M$_\odot$), above which all UCDs were found to have 
enhanced M/L, indicative of either the presence of a dark mass (in this case massive black holes as the baryonic densities are too high for dark matter 
to significantly affect the M/L) or an IMF change, and below which UCDs display bimodal M/L, some being consistent with normal stellar populations 
and no dark mass. This mass is also in the middle of the range suggested by \cite{Chilingarian11} who found that above 10$^8$ M$_\odot$, tidally 
threshed objects were dominant, while below 10$^7$ M$_\odot$ objects associated with red (metal-rich) GC formation were the norm.

To these previous findings we add the observation from Figure \ref{fig:Mv_Reff_Hist} that the number of compact stellar systems (with R$_{\rm e}$ $<$ 
400pc) appears to drop notably, to an almost constant value above M$_{\rm V}$ $\sim$ --13. As previously discussed, although this finding relies on a 
heterogeneous dataset, it is also the case that the brighter objects should be easier to find, so we doubt that any simple selection effect can be affecting 
all previous studies to create this feature.

\begin{figure} 
   \centering
   \begin{turn}{0}
   \includegraphics[scale=0.625]{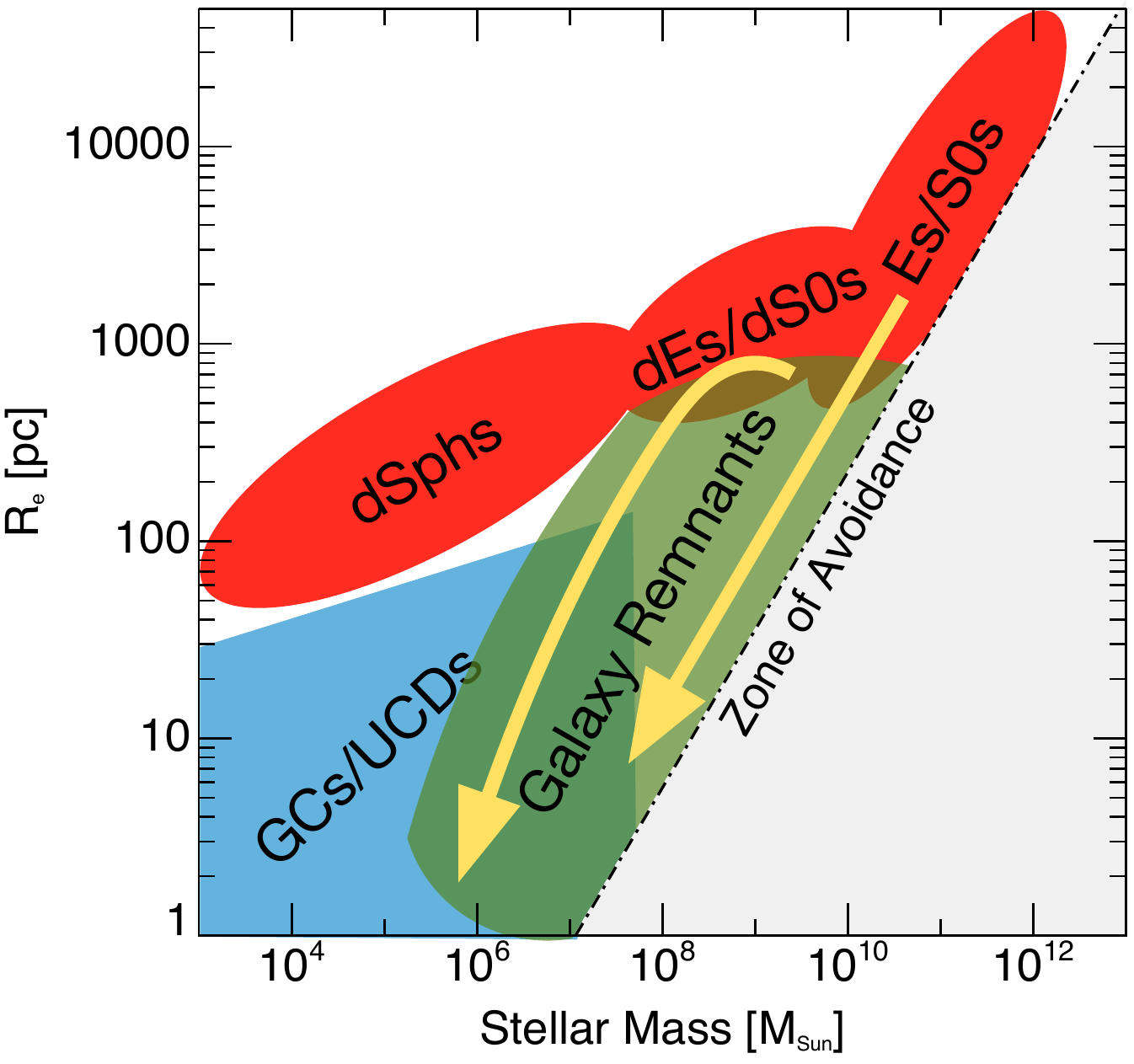}
   \end{turn} 
   \caption{Idealised version of Figure \ref{fig:Mass_Reff} showing the location of different compact stellar systems and galaxy reference samples in the 
   mass--size plane. The red ellipses show the location of the various early-type galaxy sequences, and the blue wedge shows the location of star cluster type 
   systems (i.e. GCs and UCDs) including their upper mass limit at around 7$\times$10$^{\rm 7}$ M$_{\odot}$. The green region shows the location of objects 
   formed through the stripping of larger galaxies, which have previously been called cEs (if stripped bulges) or UCDs (if stripped dwarf galaxy nuclei) but are 
   really members of the same sequence of objects. The yellow arrows show idealised evolutionary tracks for galaxies being stripped, with the left track being 
   nucleated dwarfs undergoing stripping, the right track is for bulged Es, S0s, and spirals. Additionally, some cEs could also represent the extension of the 
   classical E sequence to low mass.}
   \label{fig:cartoon}
\end{figure}

Taking these observations together, it is possible to sketch out a rough outline of where the different stellar systems lie in the mass--size plane (Figure
\ref{fig:cartoon}). The red ellipses show the regions inhabited by ellipticals, early-type dwarfs, and dSphs; and the blue region shows where star cluster 
type  objects can be found, including their upper mass limit at $\sim$7$\times$10$^{7}$ M$_{\odot}$. It should be noted that even star clusters may 
themselves have multiple distinct origins \citep[e.g. ][]{Elmegreen08, Pfalzner09,Baumgardt10}, although this would not affect our conclusions about the 
formation of other stellar systems. The green region indicates where the stripped remains of galaxies are expected to be found, based on known objects 
which are strongly suspected to be stripped: several cEs and M~32, the brightest known literature UCDs, Perseus-UCD13 and NGC~4546-UCD1, and 
the known locations of dwarf nuclei. The yellow arrows are illustrative tracks for objects being stripped, with the left track for nucleated dwarf galaxies 
and the right for bulged massive galaxies. It is also possible that some fraction of the more massive objects (cEs in general) are a low mass extension
of the classical elliptical sequence as suggested by \cite{Kormendy12}.

\subsection{Predictions for Other Compact Stellar System Properties}

In this paper we have described the luminosity, size, stellar mass, and velocity dispersion behaviour of compact stellar systems. Given the conceptual framework 
for CSS formation suggested in Figure \ref{fig:cartoon} we can make several predictions for other CSS properties.

The first prediction is that star clusters vs. stripped (or giant elliptical sequence extension) CSS properties should display bimodality. For example, stripped nuclei 
or bulges are expected to display multi-component surface brightness profiles, due to the lingering presence of remnants of the outer galaxy structures \citep[see 
e.g. ][]{Pfeffer13}. In contrast, massive star clusters are likely to retain the simple structures of their lower mass counterparts. Additionally the orbital properties of 
CSSs formed by each route are likely to be different. As discussed in \cite{Strader11b} CSSs formed by the stripping of nuclei and bulges are expected to have 
preferentially radial orbits, as they are formed by stripping during close passages of the more massive galaxy. Conversely, because of survivor bias, massive star 
clusters should be preferentially on tangential orbits as these avoid close passages of the galaxy centre.

A further prediction is that CSSs formed through stripping should contain intermediate-mass/super-massive black holes with masses appropriate for the original 
mass of the galaxy before it was stripped. This should in general mean that they are over massive relative to the standard M$_{\rm BH}$--M$_\star$ relation. 
In contrast star cluster type CSSs should either have no massive black holes or at most ones that follow the standard M$_{\rm BH}$--M$_\star$ relation. Work 
by \citet{Mieske13} has already demonstrated that this seems to be the case; they find that massive CSSs have significant amounts of dark mass consistent with 
over massive central black holes, while lower mass CSSs hint at a bimodal distribution of dark mass with some consistent with no dark mass (presumably star cluster 
type CSSs) and some consistent with significant dark mass (presumably stripped objects).

CSSs are also expected to differ in their stellar populations. The majority of the GC populations of galaxies are ancient (with age $\sim$ 10 Gyr), whereas stripped 
nuclei and bulges can potentially display a range of ages, especially in the field, where actively star forming galaxies can be stripped. Therefore spectroscopically 
determined ages are a potentially powerful way to separate star clusters from stripped nuclei in the CSS population. There are also likely to be environmental 
dependences of the CSS stellar populations, in the sense that in lower density environments the galaxies being stripped are more likely to be later-type and hence 
younger than the objects being stripped in denser environments (primarily dEs and dS0s), leading to the prediction that field/group stripped objects should on average 
be younger than cluster stripped CSSs. 

Metallicities of stripped nuclei and bulges should be significantly higher than expected for their current masses, if they were really merely an extension of the
early-type galaxy population. This is because early-type galaxies display a mass--metallicity relation, in the sense that more massive galaxies are more metal rich 
(up to a point), while stripping reduces mass without affecting the central metallicity of the stellar population \citep{Chilingarian09}. Recent work (by e.g. 
\citealt{Chilingarian09,Francis12}) already demonstrates that massive cEs including M32 and NGC4486B have metallicities consistent with those of much larger 
galaxies, rather than with the observed mass--metallicity relation for early-types. In fact M32 is offset from the normal mass--metallicity relation by more than 
0.6 dex in metallicity (or alternatively by more than 4 magnitudes in luminosity). In the case where there is an additional population of true low mass classical 
ellipticals the predictions would be generally the same as those for the stripped nuclei with the possible exception of the metallicities. The metallicities of 
low mass classical ellipticals could follow the general mass-metallicity trend, alternatively, the high density of the objects (leading to a higher potential) could
lead to higher levels of enrichment than less compact early types of the same mass due to the increased retention of supernova ejecta.

\subsection{The Zone of Avoidance}
\label{Sec:ZoA}

Figure \ref{fig:Mass_MassSurf} shows an apparent mass dependence of the maximum of the effective surface mass density varying from $\sim$10$^{6}$ M$_{\odot}$ 
pc$^{-2}$ for galaxy nuclei, to $\sim$10$^{3}$ M$_{\odot}$ pc$^{-2}$ for the most massive ellipticals. However, \citet{Hopkins10} demonstrate that in fact nuclei, the 
MW Nuclear Disk, Cen A GCs, UCDs, and ellipticals all reach the same maximum surface mass density $\Sigma_{\rm max}$ of $\sim$10$^{5}$ M$_{\odot}$ pc$^{-2}$. 
The apparent mass trend in effective surface mass density $\Sigma_{\rm eff}$ is simply the result of the evolutionary change in the structures of the objects leading to 
the effective radius being larger, and the effect of the central structure being diluted. In the simplest example the $\Sigma_{\rm max}$ of a nucleated dwarf clearly 
must be roughly the same as that of a pure nucleus, but the $\Sigma_{\rm eff}$ will be considerably lower, because the exponential component of the dwarf has a lower 
mass surface density and extends the effective radius of the structure considerably. \citet{Hopkins10} also noted that the maximum surface mass density is reached over 
a considerable range of scales for the different objects, indicating that it is not simply a limit on the three-dimensional stellar density.

\citet{Hopkins10} ascribed the observation of a constant $\Sigma_{\rm max}$ for nuclei, UCDs, and ellipticals to feedback from massive stars in the baryon-dominated 
cores of star clusters and galaxies. One potential problem with this picture is that galaxy nuclei do not appear to form in single formation events \citep[see e.g.][]{Walcher06,Seth10}, 
and if they are gradually built up over time in a series of smaller formation events, this feedback is likely to prove less effective as the radiation field from young massive
stars in each star formation event is smaller than would be present if all of the mass was produced simultaneously

Nevertheless, although we have searched for objects that breach the observed effective surface mass density limit we have not yet found any culprits. It therefore appears 
that at least one physical process is responsible for limiting the $\Sigma_{\rm max}$ and through a conspiracy with the structural changes of these various object types
this leads to a well defined ``zone of avoidance". Whatever this physical process is, it is responsible for limiting $\Sigma_{\rm max}$ for a diverse group of objects that 
had a wide range of formation processes: from formation in a one-off near instantaneous burst (the GCs), through repeated smaller star formation events (the nuclei), up 
to building up through hierarchical merging over Gyr timescales (the ellipticals).

\section{Conclusions}

We have presented the first results from the AIMSS survey. By examining the luminosities, masses, effective radii 
and velocity dispersions of a sample of newly discovered compact stellar systems, along with literature compilations,  
we have reached the following conclusions:

\begin{enumerate}

\item We have discovered several new stellar systems which completely bridge the gap between star
clusters and galaxies in the mass--size, mass--mass surface density, and mass--velocity dispersion planes.
	
\item Three of our newly discovered compact stellar systems (NGC~1128-AIMSS1, NGC~1132-UCD1, 
ESO~383-G076-AIMSS1) are the closest known M32 analogues found to date. These objects are significantly
more massive than typical UCDs of similar radius. When combined with three other unusually dense stellar
systems and M32, their environmental distribution shows that these objects can be formed in all galactic 
environments from the field to galaxy clusters. The relative rarity of these objects, combined with their 
environmental ubiquity, might point to them being formed often but being short lived when they do arise.

\item The existence of our compact stellar systems, along with other recently discovered objects, throws 
into doubt a universal well-defined mass--size relation for compact stellar systems. These objects do however, 
further support the existence of a universal ``zone of avoidance" for all dynamically hot stellar systems, 
beyond which no isolated system can add stellar mass at fixed size. 

\item By examining the luminosity distribution of compact stellar systems, we offer further support for the idea
that M$_{\rm V}$ $\sim$ --13 (M$_\star$ $\sim$ 7 $\times$ 10$^7$ M$_{\odot}$) is a fundamental limit for the 
creation of compact stellar systems in star cluster formation processes. At larger masses all objects, whether 
called UCDs or cEs, are likely created by the tidal stripping of larger galaxies, though we cannot rule out 
dissipative merging in some cases. Below this mass a combination of star clusters and stripped nuclei exist, 
with the fraction of star clusters increasing towards lower mass.

\item We suggest that two types of UCD/cE-like object exist, one type being the result of the tidal stripping of 
galaxies with bulges (Es, S0s, and spirals), and the other the result of the stripping of nucleated dwarfs
(dEs/dS0s).

\item Finally, the fact that our compact stellar systems are found associated with galaxies 
located in a range of environments from the field/loose groups to the densest clusters indicates that while dense environments 
may aid compact stellar system formation (especially for the most massive CSSs which come from larger 
galaxies), they are not essential.

\end{enumerate}

\section{Acknowledgements}

We would like to thank Joel Pfeffer for providing the output of several of his dE, N stripping simulations.
We would also like to thank Remco van den Bosch and Eva Schinnerer for helpful discussions 
and comments, as well as Bart Dunlap and Brad Barlow for carrying out some of the observations
this paper is based upon.

Support for Program number HST-AR-12147.01-A was provided by NASA through a grant 
from the Space Telescope Science Institute, which is operated by the Association of Universities 
for Research in Astronomy, Incorporated, under NASA contract NAS5-26555.

DAF thanks the ARC for financial support via DP130100388. This work was supported by National
Science Foundation grant AST-1109878.

SJP acknowledges the support of an Australian Research Council Super Science Postdoctoral Fellowship 
grant FS110200047.

FRF and AVSC acknowledge financial support from Consejo Nacional de Investigaciones Cient\'ificas 
y T\'ecnicas, Agencia Nacional de Promoci\'on Cient\'ifica y Tecnol\'ogica (PICT 2010-0410), and Universidad 
Nacional de La Plata (Argentina).

Based on observations obtained at the Southern Astrophysical Research (SOAR) telescope, 
which is a joint project of the Minist\'{e}rio da Ci\^{e}ncia, Tecnologia, e Inova\c{c}\~{a}o (MCTI) 
da Rep\'{u}blica Federativa do Brasil, the U.S. National Optical Astronomy Observatory (NOAO), 
the University of North Carolina at Chapel Hill (UNC), and Michigan State University (MSU).

Some of the observations reported in this paper were obtained with the Southern African Large 
Telescope (SALT).

Some of the data presented herein were obtained at the W.M. Keck Observatory, which is 
operated as a scientific partnership among the California Institute of Technology, the University 
of California and the National Aeronautics and Space Administration. The Observatory was 
made possible by the generous financial support of the W.M. Keck  Foundation.

The authors wish to recognize and acknowledge the very significant cultural role and reverence 
that the summit of Mauna Kea has always had within the indigenous Hawaiian community.  We 
are most fortunate to have the opportunity to conduct observations from this mountain.

This paper makes use of data obtained as part of Gemini Observatory programs  GS-2004A-Q-9 
and GS-2011A-Q-13. 	

Based on observations obtained at the Gemini Observatory, which is operated by the 
Association of Universities for Research in Astronomy, Inc., under a cooperative agreement 
with the NSF on behalf of the Gemini partnership: the National Science Foundation 
(United States), the National Research Council (Canada), CONICYT (Chile), the Australian 
Research Council (Australia), Minist\'{e}rio da Ci\^{e}ncia, Tecnologia e Inova\c{c}\~{a}o 
(Brazil) and Ministerio de Ciencia, Tecnolog\'{i}a e Innovaci\'{o}n Productiva (Argentina).

Based on observations made with the Isaac Newton Telescope (INT) operated on 
the island of La Palma by the Isaac Newton Group (ING) in the Spanish Observatorio del Roque de 
los Muchachos of the Instituto de Astrofisica de Canarias (IAC).

This research has made use of the NASA/IPAC Extragalactic Database (NED) which is operated 
by the Jet Propulsion Laboratory, California Institute of Technology, under contract with the 
National Aeronautics and Space Administration.

\bibliographystyle{mn2e}
\bibliography{references}

\appendix
\section{Catalog}
\label{Sec:Appendix}

In Table \ref{tab:catalog} we provide the complete catalog of CSSs and comparison samples described in 
Section \ref{Sec:LitComSam}. The literature sample is constructed from the following sources: \\

\cite{SDSSDR9}, \cite{Bastian06}, \cite{Bastian13}, \cite{Brodie11}, \cite{ATLAS3DI}, \cite{ATLAS3DXX}, 
\cite{ATLAS3DXV}, \cite{Chiboucas11}, \cite{Chilingarian07}, \cite{Chilingarian08a}, \cite{Chilingarian08b},
\cite{Chilingarian09b}, \cite{Chilingarian10a}, \cite{Chilingarian11}, \cite{Denicolo05a}, \cite{Evstigneeva07a}, \cite{Evstigneeva08},
\cite{Firth07}, \cite{Forbes11}, \cite{Forbes13}, \cite{Geha02}, \cite{Geha03}, \cite{Goudfrooij01}, 
\cite{Gregg09}, \cite{Harris96}, \cite{Hau09}, \cite{Huxor11b}, \cite{Huxor13}, \cite{Jones06}, \cite{Karick03},
\cite{Madrid11}, \cite{Madrid13}, \cite{Maraston04}, \cite{McConnachie12}, \cite{Misgeld&Hilker11}, 
\cite{Mieske04}, \cite{Mieske06}, \cite{Mieske08b}, \cite{Mieske08}, \cite{Mieske08c}, \cite{Mieske13},
\cite{Norris&Kannappan11}, \cite{Norris12}, Penny et al. (in prep), \cite{Pota13a}, \cite{Price09},
\cite{Rejkuba07}, \cite{Schweizer&Seitzer98}, \cite{Schweizer07}, \cite{Smith-Castelli13}, \cite{Strader11a},
\cite{Strader13},\cite{Taylor10}, \cite{Toloba12}.

\begin{table*}
\centering
\small
\begin{tabular}{|llllllllll|}
\hline
Name 			& Type	&	RA		&	Dec			&	M$_{\rm V}$	&	M$_{\star}$		&	R$_{e}$		& $\sigma$			& References					\\
				& 		& (J2000)		&	(J2000)		&	(mag)		&	(M$_\odot$)		&	(pc)			& (km\,s$^{-1}$)		&							\\
\hline
NGC 0315		& 1		& 14.453681   	&	30.352448	& --24.6			&  2.704$\times10^{12}$	&	30619.6		&	351.6			& Misgeld $\&$ Hilker 2011		\\
NGC 0584		& 1		& 22.836479   	&	-6.868061		& --22.6			&  3.428$\times10^{11}$	&	5296.6		&	217.3			& Misgeld $\&$ Hilker 2011		\\
NGC 0636		& 1		& 24.777204     &	-7.512603		& --21.6			&  1.225$\times10^{11}$	&	3647.5		&	156.3			& Misgeld $\&$ Hilker 2011		\\
...				& ...		& ...		   	&	...			& ...				&  ...					&	...			&	...				& ...							\\

\hline
\end{tabular}

\caption[Catalog of CSSs and comparison samples]{Catalog of CSSs and comparison samples. The first three objects in the catalog are
shown to demonstrate the form of the catalog, all objects are available in the electronic version of this table. The columns are Name, Type, 
R.A., Dec., absolute V band magnitude, stellar mass, effective radius, $\sigma$, and literature reference for catalog objects. The type codes refer to the object types 
as follows 1) Es/S0s, 2) dEs/dS0s, 3) dSphs, 4) Nuclear Star Clusters, 5) Literature GCs, UCDs, cEs, 6) AIMSS GCs, UCDs, cEs, 7) YMCs.
Stellar masses are from the literature sources listed except for types 2, 5, 6, and 7 where the stellar masses were computed following the
approach outlined in Section \ref{Sec:StellMass}.

}
\label{tab:catalog} 
\end{table*}

\label{lastpage}

\end{document}